\documentclass[preprint]{elsarticle}

\usepackage{amsmath}
\usepackage{amssymb}
\usepackage{amsthm}

\usepackage[T1]{fontenc}
\usepackage[utf8]{inputenc}
\usepackage{tikz}
\usepackage{pgfplots}
\usepackage{algorithm}
\usepackage[noend]{algorithmic}

\usepackage{enumitem}
\usepackage{amsfonts, amssymb, amsmath}
\usepackage{mathtools}
\usepackage{booktabs}
\usepackage{tabularx}
\usepackage{fullpage}
\usepackage{multirow}
\usepackage{url}
\usepackage{bbold}
\usepackage{grffile}
\usepackage{appendix}
\usepackage{subcaption}

\newtheorem{theorem}{Theorem}[section]

\newtheorem{proposition}[theorem]{Proposition}

\DeclareMathOperator*{\argmin}{arg\,min}

\begin{document}

\begin{frontmatter}

\title{Heuristic Approaches to Obtain Low-discrepancy Point Sets via Subset Selection}
\author[1]{François Clément}
\ead{francois.clement@lip6.fr}
\author[1]{Carola Doerr}
  \ead{Carola.Doerr@lip6.fr}
  \author[2]{Lu\'is Paquete}
  \ead{paquete@dei.uc.pt}

  \address[1]{Sorbonne Universit\'e, CNRS, LIP6, Paris, France}
  \address[2]{University of Coimbra, CISUC, Department of Informatics Engineering, Portugal}

\begin{abstract}

Building upon the exact methods presented in our earlier work [J. Complexity, 2022], we introduce a heuristic approach for the star discrepancy subset selection problem. The heuristic gradually improves the current-best subset by replacing one of its elements at a time. While the heuristic does not necessarily return an optimal solution, we obtain very promising results for all tested dimensions. For example, for moderate 
sizes $30 \leq n \leq 240$, we obtain point sets in dimension 6 with 
$L_{\infty}$ star discrepancy up to 35\% better than that of the first $n$ points of the Sobol' sequence. Our heuristic works in all dimensions, the main limitation being the precision of the discrepancy calculation algorithms.

We provide a comparison with a recent energy functional introduced by Steinerberger [J. Complexity, 2019], showing that our heuristic performs better on all tested instances. Finally, our results and complementary experiments also give further empirical information on inverse star discrepancy conjectures.
\end{abstract}


\end{frontmatter}
\section{Introduction}

Discrepancy measures are designed to quantify how regularly a point set is distributed in a given space. Many discrepancy measures exist, one of the most common ones being the $L_{\infty}$ star discrepancy. The $L_{\infty}$ star discrepancy of a point set $P \subseteq [0,1]^d$ measures the worst absolute difference between the Lebesgue measure of a $d$-dimensional box anchored in $(0,\ldots,0)$ and the proportion of points that fall inside this box. The $L_{\infty}$ star discrepancy gained significant attention because of the Koksma-Hlawka inequality~\cite{Hlawka,Koksma} which bounds the error made in numerical integration. While Quasi-Monte Carlo integration is their main application~\cite{DickP10}, point sets of low $L_{\infty}$ star discrepancy are also used for one-shot optimization~\cite{CauwetCDLRRTTU20}, design of experiments~\cite{SantnerDoE}, computer vision~\cite{MatBuilder}, and financial mathematics~\cite{GalFin}.

Throughout this paper, we write $n$ for the number of points of a $d$-dimensional point set. We write \emph{sequence} for an infinite series of points and \emph{set} for a finite one, both are closely linked as results on sequences in dimension $d$ correspond to those on sets in dimension $d+1$~\cite{Roth54}. Well-known constructions such as Sobol', Halton, Hammersley, or digital nets achieve discrepancies of order $\log^{d-1}(n)/n$ for \emph{sets} of $n$ elements in dimension $d$ (see~\cite{Nie92} for a detailed description). This becomes $\log^{d}(n)/n$ for a fixed \emph{sequence} for which we consider all possible $n$, see~\cite{Roth54} for details.  
However, the driving focus behind the construction of these sequences is to obtain the optimal \emph{asymptotic} order for the $L_{\infty}$ star discrepancy. While random points -- whose star discrepancy scales in $\sqrt{d/n}$~\cite{Doerr14lowerBoundRandomPoints} -- are not as good in the asymptotic setting, they often outperform low-discrepancy sequences for practical applications where the number of samples $n$ might be too small to reach the asymptotic advantage~\cite{Bergstra}.

The lack of low-discrepancy constructions adapted to specific $n$ and $d$ combinations led us to introduce the Star Discrepancy Subset Selection Problem (SDSSP) in~\cite{CDP}. We showed that the SDSSP is an NP-hard problem and proposed two exact methods, a branch-and-bound algorithm and an MILP formulation, to solve it for low $n$ and $d \in \{2,3\}$. In this specific setting, we obtained point sets of much better discrepancy than the known low-discrepancy sets of the same size (Sobol', Halton, and Faure for example), with only the Fibonacci sequence being competitive for specific point set sizes in dimension 2.

\textbf{Our contribution}: Extending~\cite{CDP}, we provide in this work a heuristic to solve the problem in much higher dimensions, as well as with a higher number of points. We introduce a swap-based heuristic, which attempts to replace a point of the chosen subset by one currently not chosen, using the box with the worst local discrepancy to guide our swap-choice. A further brute-force check is then used to guarantee that the final point set is a local minimum. This approach is able to obtain point sets of much smaller discrepancy than the known constructions for all dimensions for which the discrepancy can be reliably computed, hence significantly improving on the range of settings that can be handled by the exact methods presented in~\cite{CDP}.

With different instantiations of the heuristic, we obtain point sets that are between 10 and 40\% better than the initial Sobol' set of the same size. Initial experiments also show that choosing a subset of size $k$ close to the initial point set size~$n$ leads to better results, especially when the dimension $d$ and the target set size~$n$ increase, which was also observed with exact methods.

We also compare our method with the energy functional on the points' position introduced by Steinerberger~\cite{StefEnerg}, which is minimized by gradient descent to obtain a point set with low discrepancy. His approach can be used on any point set in any dimension but he provides results mostly for dimensions 2 and 3. We give a detailed comparison of his method with ours as well as some extended testing in higher dimensions in Section~\ref{CompaSte}. We show that not only can we clearly outperform the results obtained by this approach, but also combining the two methods allows us to build point sets whose discrepancy is competitive with that of the Sobol' sequence. This can be done with any starting point set, it does not require a good number-theoretic construction. We note that the sets obtained this way are not as good as those obtained by using subset selection on the Sobol' sequence. 

Finally, our experiments provide numerous discrepancy values for the Sobol' sequence for varying $n$ and~$d$. It is conjectured in~\cite{NW} that $n=10d$ points are required to obtain a discrepancy of $0.25$ in dimension $d$. Our results in Section~\ref{sec:inverse} show that the Sobol' sequence seems to come close to a discrepancy of $0.2$ with $n=10d$ points in all tested dimensions. Furthermore, our improved sets obtained via subset selection come closer to $n=7d$ points required to reach a discrepancy of $0.25$.

\textbf{Related work}: Despite the extensive research on optimal asymptotic constructions, there is little similar work attempting to build point sets directly. Apart from the mentioned work by Steinerberger~\cite{StefEnerg}, the most similar work was by Doerr and de Rainville in~\cite{Rainville} where they optimize, using evolutionary algorithms, the permutations used to build generalized Halton sequences. Their work had a very similar goal but was limited to improving one specific sequence.

\textbf{Structure of the paper:} We recall in Section~\ref{sec:back} some important results on the $L_{\infty}$ star discrepancy. In particular, we give a brief reminder on the structure of the $L_{\infty}$ star discrepancy function, describe the two main methods of computing it and recall the main results on the Star Discrepancy Subset Selection problem. In Section~\ref{sec:heu} we introduce the new heuristics to solve the problem in higher dimensions. Section~\ref{sec:results} will describe our numerical evaluation and the quality of the obtained point sets, as well as a comparison with Steinerberger's energy functional.

\textbf{Availability of code}: Our code and obtained point sets are available at \url{https://github.com/frclement/SDSSP_Heuristics}.

\section{The $L_{\infty}$ Star Discrepancy}\label{sec:back}
\subsection{General Results on the Star Discrepancy}

The $L_{\infty}$ star discrepancy of a point set represents the worst absolute difference between the proportion of points that fall inside a box and the proportion of volume taken by this box. Formally, the $L_{\infty}$ star discrepancy of a point set $P\subseteq [0,1]^d$ is defined as

\begin{equation}
	d_{\infty}^*(P) := \sup_{q \in [0,1]^d} \left \lvert \frac{|P \cap [0,q)|}{|P|}   
		- \lambda(q) \right \rvert, 
\label{eq:1}
\end{equation}
where $\lambda(q)$ is the Lebesgue measure of the box $[0,q)$.

For the asymptotic order of the $L_{\infty}$ star discrepancy, there exist several sequences reaching an order of $O(\log^{d}(n)/n)$. These are known as \emph{low-discrepancy sequences} and exhibit the best \emph{asymptotic} $L_{\infty}$ star discrepancy known today. However, the best lower bound for the star discrepancy in dimension $d$ states that there exist constants $c,C$ depending only on the dimension such that $d^{*}_{\infty}(P) \geq c\log^{C+(d-1)/2}(n)/n$~\cite{BilykSmall}. It is conjectured that, in any dimension $d$, there exists a constant $c_d$ such that $d^{*}_{\infty}(P) \geq c_d\log^{d-1}(n)/n$ for any point set $P\subseteq [0,1]^d$ of size $n$. This would give us a matching bound to the low-discrepancy sets mentioned previously. The difference with the $O(\log^d(n)/n)$ bound is due to the set/sequence distinction. However, this conjecture has only been proven to hold in dimensions 1 and 2~\cite{SchmidtLowerD2}. More detailed bounds can be found in~\cite{DickP10}.

Despite being defined as a continuous problem over all possible anchored boxes, calculating the star discrepancy can be treated as a discrete problem~\cite{NieBox}. First, we notice that any closed anchored box in $[0,1]^d$ can be obtained as the limit of a sequence of bigger open boxes that contain the same number of points. The only exception is $[1,\ldots,1]$ and this closed box cannot give the worst discrepancy value as its local discrepancy is 0.
Following the notation introduced in~\cite{CDP}, we define $D(q,P)$ to be the number of points of $P$ that fall inside the open anchored box $[0,q)$ and $\overline{D}(q,P)$ the number of points of P that fall inside the closed anchored box $[0,q]$. We define the two following functions:
\begin{equation}
	\delta(q, P) := \lambda(q) - \frac{1}{n} D(q, P)
\qquad \text{ and } 
\qquad 
	\overline{\delta}(q, P) := \frac{1}{n} \overline{D}(q, P) - \lambda(q).
	\label{eq:delta}
\end{equation}
The local discrepancy in a point $q \in [0,1]^d$ is given by the maximum of $\delta(q,P)$ and $\overline{\delta}(q,P)$. It is also not necessary to consider all possible values for $q$. Indeed, we can define for all $j \in \{1,\ldots,d\}$ the grid
\begin{equation}
	\Gamma(P) := \Gamma_1(P) \times \ldots \times \Gamma_d(P)
\qquad \text{ and } 
\qquad 
	\overline{\Gamma}(P) := \overline{\Gamma}_1(P) \times \ldots \times \overline{\Gamma}_d(P),
\end{equation}
with
\begin{equation}
	\Gamma_j(P) := \{x_j^{(i)} | i \in 1,\ldots,n\} 
\qquad \text{ and } 
\qquad 
	\overline{\Gamma}_j(P) := \Gamma_j(P) \cup \{1\},
\end{equation}

As shown in more detail in~\cite{DGWBook}, the star discrepancy computation reduces to the following discrete problem
\begin{equation}
d_{\infty}^{*}(P) = \max \left\{\max_{q \in \overline{\Gamma}(P)}\delta(q, P),
\max_{q \in \Gamma(P)}\overline{\delta}(q, P)\right\}.
\label{discrepancy_formula}
\end{equation}

\subsection{Star Discrepancy: Problem-Specific Algorithms}\label{sec:algos} 
\begin{figure}
    \centering
    \includegraphics[width=0.6\textwidth]{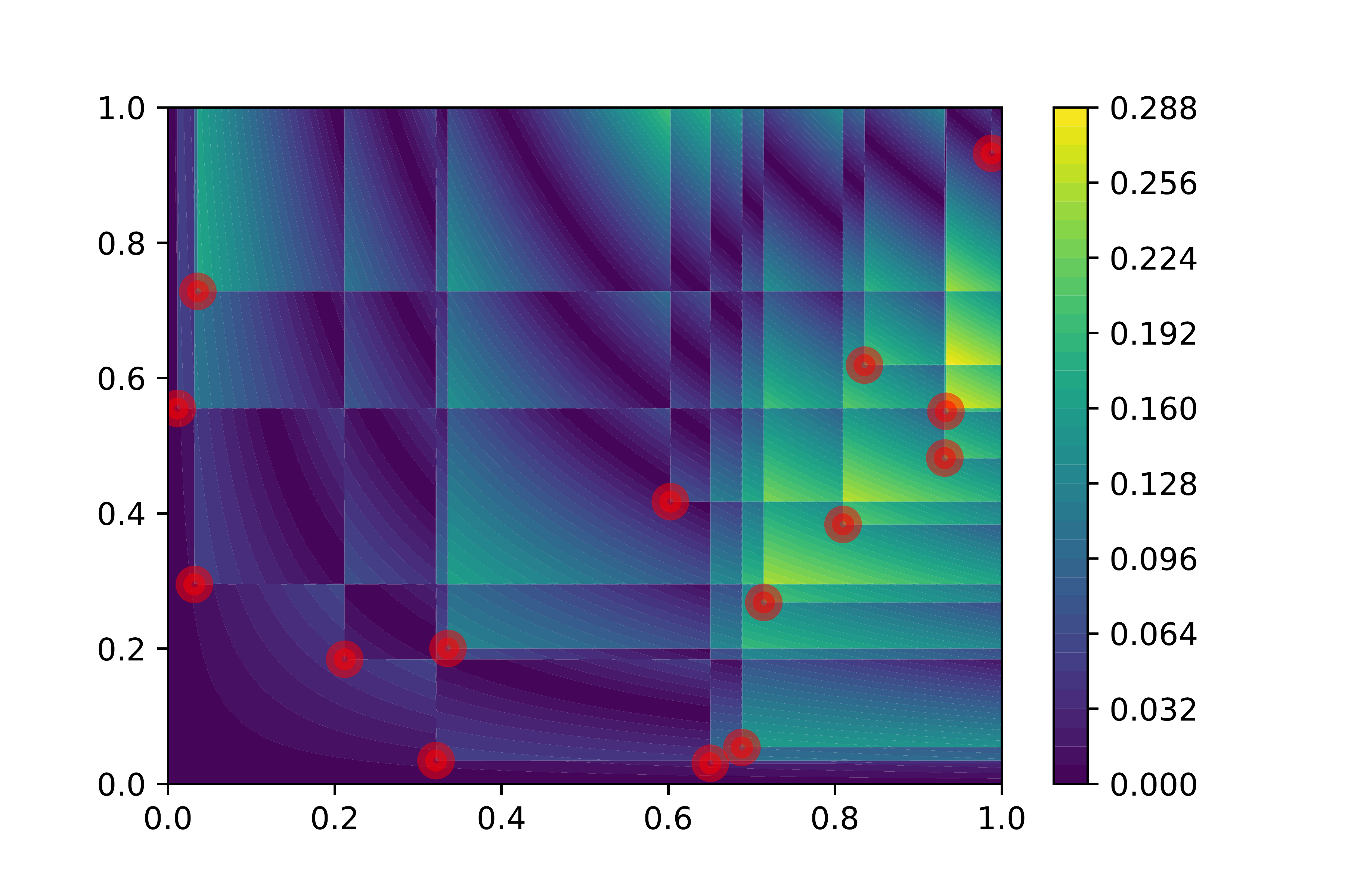}
    \caption{Local discrepancy values for a random point set of 15 points (red dots) in dimension 2}
    \label{crit}
\end{figure}

The decision version of the star discrepancy computation has been shown to be NP-hard~\cite{NPhard}, and even W[1]-hard~\cite{W1hard}. Directly solving the discrete problem introduced above would require calculating $O(n^d)$ local discrepancies, which can be reduced to $O(n^d/(d!))$ by noticing that only specific boxes can reach the maximal discrepancy value. A box $[0,q)$ (or $[0,q]$) can obtain the maximal discrepancy value only if for all $i\in \{1,\ldots,d\}$ there exists a point $p \in P$ such that $p \in [0,q]$ and $p_i=q_i$ (where we write $q=(q_1,\ldots,q_d)$). 
These specific boxes $[0,q)$ and $[0,q]$ are called \emph{critical boxes}. In Figure~\ref{crit}, the critical boxes are visible as the intersection points of the grid lines. Critical boxes are defined for both open and closed boxes, the difference being that for open boxes the points defining its edges will not be inside the box.

The best known exact algorithm to compute the star discrepancy value of a given point set, proposed by Dobkin, Eppstein and Mitchell in~\cite{DEM}, runs in time $O(n^{1+d/2})$. Based on a space decomposition originally used to solve Klee's measure problem, the algorithm carefully builds $O(n^{d/2})$ disjoint boxes covering $[0,1]^d$. The algorithm then finds the worst discrepancy in each box in linear time via dynamic programming, giving the stated complexity. We will be writing \textit{DEM algorithm} to refer to this algorithm from now on.

On the heuristics side, the best known algorithm was proposed in~\cite{GnewuchWW12}. It is built on top of an iterative search heuristic called \emph{Threshold Accepting}; we will hence  refer to this algorithm as the \emph{TA heuristic}. The algorithm walks along the grid structure, evaluates local discrepancy values of some of the grid points, and outputs the largest value seen. It therefore always outputs a lower bound for the discrepancy of the point set under consideration. Based on the results presented in~\cite{GnewuchWW12}, this bound reliably matches the true discrepancy value until somewhere between dimensions 12 and 20 for a few hundred points. For larger settings, we do not have any means to estimate the tightness of the returned lower bound. 

A more detailed description of these algorithms and other attempts to compute the star discrepancy can be found in Chapter 10 in~\cite{DGWBook}.

\subsection{The Star Discrepancy Subset Selection Problem (SDSSP)}\label{sec:SDSSP}

We recall the \emph{star discrepancy subset selection problem (SDSSP)} introduced in~\cite{CDP}. Given a point set $P \subseteq [0,1]^d$ of size $|P|=n$ and an integer $0<k \leq n$, find a subset $P^{*}$ of P such that $|P^{*}|=k$ and $d_{\infty}(P^{*})$  is minimal among all other subsets of $P$ of size $k$. This comes down to solving the problem
\begin{equation}
\argmin_{\substack{P^* \subseteq P\\|P^*| = k}} \max \left \{\max_{q \in \overline{\Gamma}(P^{*})}\delta(q, P^{*}), \max_{q \in \Gamma(P^{*})}\overline{\delta}(q, P^{*}) \right\},
\end{equation}
which is equivalent to
\begin{equation}
\argmin_{\substack{P^* \subseteq P\\|P^*| = k}} \max \left \{\max_{q \in \overline{\Gamma}(P)}\delta(q, P^{*}), \max_{q \in \Gamma(P)}\overline{\delta}(q, P^{*}) \right\}.
\end{equation}
It was shown that the decision version of this problem is NP-hard. The SDSSP problem is also non-monotonic: the best solution for $k'<k$ is not necessarily contained in the one for $k$, see~\cite{CDP} for an example in dimension~1.

We introduced in~\cite{CDP} two exact methods to solve the problem. The first is an MILP formulation. It relies on introducing two constraints for each grid point (one for the open box and one for the closed box associated to each grid point), and binary variables to represent the points. The discrepancy of the obtained subset is the objective 
to minimize. We then solve the MILP with the SCIP solver~\cite{BestuzhevaEtal2021OO}), but only for dimension 2, due to solver limitations. The solver works quite well in $2d$ for up to 140 points but is very quickly limited by the runtime as well as by the size of the file describing all the constraints. It performs better when $k$ and $n$ are close, since the linear relaxation is a tighter bound (the relaxation is integral for $k=n$). In $3d$, we are not able to solve the problem with the MILP formulation.

The second approach, based on a combinatorial branch-and-bound method, considers the points in a sorted order in one dimension and at each branching step to either accept or reject a point in the subset. At each step of the branch-and-bound, we update local discrepancy values associated with well-chosen grid points to obtain a lower bound on the final discrepancy value. The branch-and-bound algorithm also successfully solves the problem in 2 and 3 dimensions for up to 140 points, but could not finish within the 30 minutes cutoff on larger instances. 

For the evaluated instances, the MILP and the branch-and-bound return subsets of up to 50\% smaller $L_{\infty}$ star discrepancy than standard low-discrepancy sets of the same size (Sobol', Halton,...).

\section{A Heuristic Approach for the Star Discrepancy Subset Selection Problem}
\label{sec:heu}
To generalize the results of SDSSP to higher dimensions and to larger point sets, we introduce in this section a general method and several instantiations of it. The main working principle of this method is to keep the best subset found so far and, at each step, to attempt replacing some of the points inside this subset with some of the currently not chosen ones. The point set with the best discrepancy is then kept: either the initial one or the set obtained after the swaps. In case of a tie, the initial set is kept. For $j \in \{1,\ldots,n\}$, we call \emph{$j$-swap} the simultaneous replacement of $j$ points inside the set with $j$ points outside the set.

The main idea is to keep a current subset and to improve it via well-chosen 1-swaps. During a first step, we only consider the points defining the worst local discrepancy box as candidates to be removed from the chosen subset. That is, if the discrepancy of the current subset $P^*$ is attained in $q \in [0,1]^d$, up to $d$ points $x \in P^* \cap [0,q]$ with $x_i=q_i$ for some $i \in \{1,\ldots,d\}$ are considered to be replaced by a point $y \in P \setminus P^*$. If no improving 1-swap is found, we then consider all remaining 1-swaps during the second step. At any point, if an improving 1-swap is found, we go back to the beginning of 
the first step with our new point set.

We note that our current choices for the heuristic are strongly influenced by the cost of discrepancy calculations. A single run with the brute-force check can require thousands of discrepancy calculations, each with a cost of $O(k^{d/2+1})$ for the DEM algorithm. Choosing carefully which swap to try should be a key focus in designing heuristics to tackle this problem. For ease of explanation, we will only consider the case when a closed box reaches the maximal local discrepancy value, the open box case is treated very similarly. A slightly simplified pseudo-code is provided in Algorithm~\ref{alg:heuristic}.

In our experiments, to compare the performance of both discrepancy calculation methods in the context of subset selection, we will highlight which of the DEM algorithm and TA heuristic are used in the experiments.

\begin{algorithm}[t]
\begin{algorithmic}
\STATE \textbf{Input}: $P$, $d$, $n$, $k$.
\STATE Let $c[0,...,d-1]$ be the critical box table.
\STATE Let $\pi_j$ be an ordering of $P\cup\{1,\ldots,1\}$ in dimension $j$, for all $j \in \{0,\ldots,d-1\}$.
\STATE $d_{min}=\infty$, $P_{min}=\emptyset$
\FOR{$it=0$ to $nb_{iter}-1$}
\STATE Select $P_k$ randomly from $P$, $|P_k|$ = $k$
\WHILE{$P_k$ is not a local minimum}
\STATE found=False
\FOR{$i=0$ to $n-1-\min(c[0],\ldots,c[d-1])$} 
\FOR{$j=0$ to $d-1$} 
\IF{$\pi_j(c_j)+i<n$} 
\IF{$d_{\infty}^*(P_k)>d_{\infty}^*(P_k\backslash\{p_{\pi_j(c[j])}\}\cup \{p_{\pi_j(c[j]+i)}\})$}
    \STATE $P_k=P_k\backslash\{p_{\pi_j(c[j])}\}\cup \{p_{\pi_j(c[j]+i)}\}$
    \STATE Update $c[0,\ldots,d-1]$, found=True
    \STATE break
\ENDIF

\ENDIF
\ENDFOR
\IF{found=True}
\STATE break
\ENDIF

\ENDFOR
\IF{found=False}
\STATE Try all remaining swaps until one improves the set or all have been tested ($P_k$ is a local minimum).
\ENDIF
\ENDWHILE
\IF{$d_{\infty}^*(P_k)<d_{min}$}
    \STATE $d_{min}=d_{\infty}^*(P_k)$
    \STATE $P_{min}=P_k$
\ENDIF
\ENDFOR
\STATE Return($P_{min},d_{min}$)

\end{algorithmic}
\caption{Pseudocode of our heuristic subset selection strategy with $nb_{iter}$ restarts}
\label{alg:heuristic}
\end{algorithm}

\textbf{Initialization.} Let $P \subseteq [0,1]^d$ with $|P|=n$ be the input point set and $k \leq n$ the target size. For each coordinate $i \in \{1,\ldots,d\}$, let $\pi_i$ be the permutation ordering the points $P \cup {(1,\ldots,1)}$ by their $i$-th coordinate. In other words, $p_{\pi_i(1),i}\leq p_{\pi_i(2),i} \leq \ldots \leq p_{\pi_i(n+1),i}=1$.

The algorithm is initialized with a randomly selected subset $P_{1}$, for which the discrepancy $d_{\infty}^{*}(P_1)$ and the corner $a=(a_1,\ldots,a_d)$ of the closed box $B_1=[0,a]$ defining this discrepancy value are computed. To improve this subset, points in $B_1 \cap P_1$ must be replaced by points in $P \backslash (B_1 \cap P_{1})$, otherwise the local discrepancy for $B_1$ will at best stay the same. Since the maximum local discrepancy can only be reached for a critical box, in every dimension $j \in \{1,\ldots,d\}$ there exists $p_{\pi_j(c_j)}$ with $c_j \in \{1,\ldots,n+1\}$ such that $p_{\pi_j(c_j)}=a_j$ and $p_{\pi_j(c_j)} \in B_1$. These points will be called \emph{edge points}, written in the $c[0,\ldots,d-1]$ table in the code.

\textbf{Step 1: Breadth-first search.} A dimension $j \in \{1,\ldots,d\}$ is picked at random and the heuristic checks if $p_{\pi_j(c_j+1)}$ is in $P_{1}$. If it is not, we compute the discrepancy of $(P_{1} \backslash \{p_{\pi_j(c_j)}\}) \cup \{p_{\pi_j(c_j+1)}\}$. If this discrepancy is \emph{strictly} lower than that of $P_{1}$, the chosen subset is changed to $P_{2}:=(P_{1} \backslash \{p_{\pi_j(c_j)}\}) \cup \{p_{\pi_j(c_j+1)}\}$. There is a new critical box $B_2$ returned during the discrepancy computation and we go back to the beginning of Step 1. If $d_{\infty}^*(P_2) \geq d_{\infty}^*(P_1)$ or if $p_{\pi_j(c_j+1)}$ is in $P_{1}$, dimension $j+1 \mod d$ is then considered, where we do the same operations. This continues until either we have found a better subset or all dimensions have been considered. If all dimensions are checked without finding an improvement, the heuristic goes back to the first dimension considered. It then does the same operations, but with $p_{\pi_j(c_j+2)}$ rather than $p_{\pi_j(c_j+1)}$ for all $j \in \{1,\ldots,d\}$. This continues until either we find a new subset with better discrepancy or until we have tried all possible swaps between $p_{\pi_j(c_j)}$ and $p_{\pi_j(b_j)}$, with $j \in \{1,\ldots,d\}$, $b_j \in \{c_j+1,\ldots,n+1\}$, and $p_{\pi_j(b_j)} \notin P_1$. In the first case, we go back to the beginning of Step 1 with our new point subset and new worst box.

\textbf{Step 2: Brute-force check.} If none of the swaps were successful, the heuristic tries all remaining valid swaps until either a better subset is found (in which case we go back to Step 1) or until we can guarantee that our current point set is a local minimum. The valid swaps to check are of two types. They can either involve an edge point: for any $j \in \{1,\ldots,d\}$, $p_{\pi_j(c_j)}$ is swapped with $p_{\pi_j(b_j)}$ for any $b_j$ such that $b_j \in \{1,\ldots,c_{j}-1\}$ and $p_{\pi_j(b_j)} \notin P_1$. Or they swap a point strictly inside the box with one outside:  $p_{\pi_j(a_j)}$ such that $a_j \in \{1,\ldots,c_{j}-1\}$ and $p_{\pi_j(a_j)} \in P_1$ is swapped with $p_h \notin P_1 \cup B_1$ with $h \in \{1,\ldots,n\}$.

\textbf{Restarts.} Multiple runs of the heuristic with different starting positions are performed, to limit the influence of the initial subset.

\subsection{Variants of the Algorithm}
Rather than the current breadth-first search, we also tried a \textit{depth-first} search where all swaps involving $p_{\pi_j(c_j)}$ for a given $j$ were done before moving to the next dimension. This did not give any noticeable improvements, and we expect our current method to perform better as it should swap points that are on average closer, with less risk of unbalancing our subset. Efficiently finding the optimal swap and the correct number of swaps to perform at once are both open questions.

Another possible modification of our algorithm would be to allow changes in the current set if the new discrepancy is equal to, and not only strictly smaller, than the current one (``\textit{plateau moves}''). However, this less demanding selection strategy results in much worse empirical performance for our settings. We attribute this performance loss to the fact that one can keep the same discrepancy value while breaking the structure in other areas of the point set (for example, moving two points inside the worst box). This leads to the algorithm making the general structure of the point set worse while not changing the overall discrepancy value but blocking future improvements. These issues could possibly be mitigated by considering other information for tie-breaking, but we did not experiment with such ideas. 

\begin{figure}
\centering
  \includegraphics[width=0.5\textwidth]{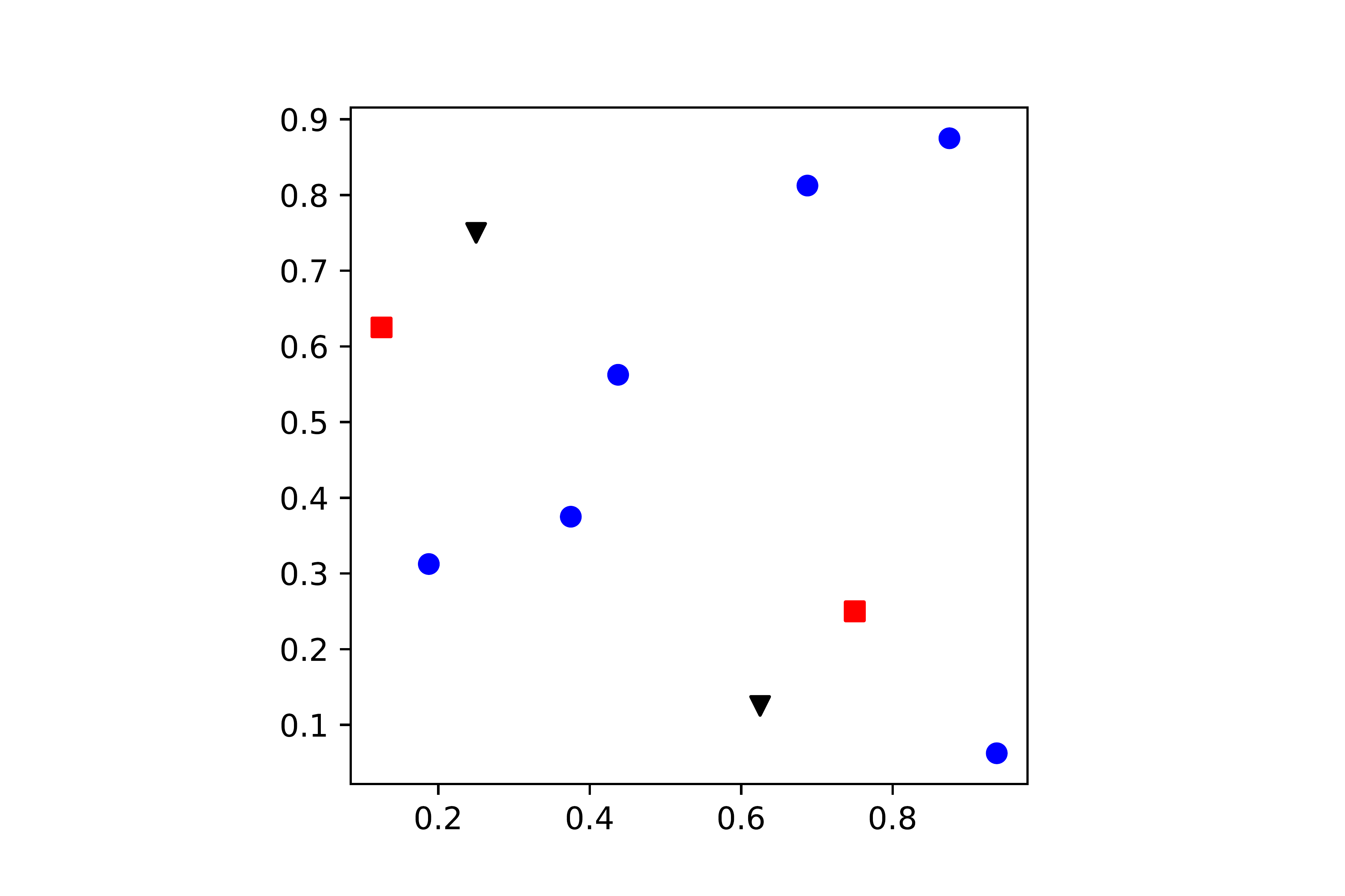}
  \caption{Two subsets for $k=8$ taken from the first $n=10$ points of the Sobol' sequence in dimension 2. The ten initial points are shown, with those present in both subsets shown as blue circles. The points shown as red squares and the blue points form a local optimum for 1-swap with discrepancy 0.234. The black triangles plus the blue points correspond to the global optimum of discrepancy 0.203. Neither of the two sets can be improved via 1-swaps. Intuitively, replacing the lower red square by the lower black triangle would create an overfilled box at the bottom, whereas replacing it by the upper triangle would create an overfilled box on the left.}
  \label{LocSub}
\end{figure}

\textbf{Non-guaranteed Optimality of Our Heuristic:}
While Section~\ref{sec:Expe} will show the algorithm's promising performance, we note that we cannot have any theoretical guarantee for the optimality of the returned set. As the proposition in~\ref{App:heur} shows, it is possible to return a local optimum that is not a global optimum. 
While the example shown in the proof is quite specific, such examples appear even for low $n$ in dimension 2 in our experiments. An example is illustrated in Figure~\ref{LocSub}.

\section{Experimental Study}\label{sec:results}
\label{sec:Expe}

In this section, we study the performance of our heuristic in different dimensions and slightly different instantiations. We also consider how the performance evolves when two of the problem's parameters are fixed, for example $d$ and $k$ or $d$ and $n$. All experiments are done on the Sobol' sequence as it gave the best discrepancy values before using subset selection. Our methods nevertheless work on any sequence or set, they only depend on the quality of the initial set to provide good results (hence our choice of Sobol'). Indeed, subset selection on random sets provides sets with much better discrepancy than the initial random set, but not as good as known low-discrepancy sets and sequences (see Section~\ref{CompaSte}). 

We also use our discrepancy calculations to provide more empirical evidence regarding conjectures on the \textit{inverse star discrepancy,} i.e., the number of points required in a given dimension to obtain a set of discrepancy less than a given threshold. We show in Section~\ref{sec:inverse} that $n=10d$ points seem to be sufficient to reach a discrepancy of 0.2. We conjecture that $n=7d$ is a closer estimate to the inverse star discrepancy of
0.25. Finally, we provide in Section~\ref{CompaSte} a detailed comparison with Steinerberger's energy functional and a potential application of the combination of the two methods.

\subsection{Experimental Setup}\label{sec:expeset}

All the different parts of the code were done in C. The Sobol' sequence generation was done with the GNU Scientific Library, using the procedure {\tt gsl\_qrng\_sobol}. Whenever randomness was required (for the first chosen subset and the dimension choice in Step 1 of the heuristics), {\tt rand()} was used and initialized with {\tt srand(1)}. The heuristics were implemented by us, with the exception of the two methods for calculating the star discrepancy (DEM algorithm and TA heuristic) which were provided to us by Magnus Wahlstr\"om and are available at \cite{reproducibiliyt}.

The experiments were run on a Debian/GNU Linux 11 computer, with a Quad Core Intel Core i7-6700 processor and 32 GB RAM. {\tt gcc} 10.2.1 was used with the {\tt -O3} compilation flag.
Experiments were run with four types of heuristic instantiations, either with the TA heuristic or the DEM algorithm, each with or without the brute-force check. The instantiations will be referred to as TA\_BF, TA\_NBF, DEM\_BF or DEM\_NBF, starting with TA if it is the TA heuristic and ending with BF if it did the brute-force check, NBF otherwise. Discrepancy values from DEM instantiations are exact but those from the TA ones are only lower bounds. They should nevertheless be relatively reliable below dimension 10.

We considered points in dimensions $\{4,5,6,8,10,15,25\}$. Unless specified otherwise, all ground sets are taken from the Sobol' sequence. Initial experiments were run for $n \in \{100,150,200,250 \}$ and $k \in \{n-10, n-20, n-30, n-40, n-50, n-60, n-70\}$, with a maximum of 10 runs for each instance. Further experiments to refine the parameter choices or get more precise results were done with slightly different values of $n$ and $k$ (but still of the same order of magnitude). They will be described for the relevant results. Each heuristic experiment was given a 1 hour cutoff, with the best value found so far returned if the heuristic had not finished by then. More precisely, if all 10 runs could finish in one hour then the value returned is the best of those, but for the larger instances the value returned may be the best value found in the first unfinished run.

\subsection{Experiment Results}\label{sec:Experes}
We describe here our experimental results, from general tests to have a global view on the performance of the heuristic, to finding the optimal parameters. The end of the section includes a brief discussion on comparisons with both sets and random subset selection.

\begin{figure}
\centering
\includegraphics[width=0.24\textwidth]{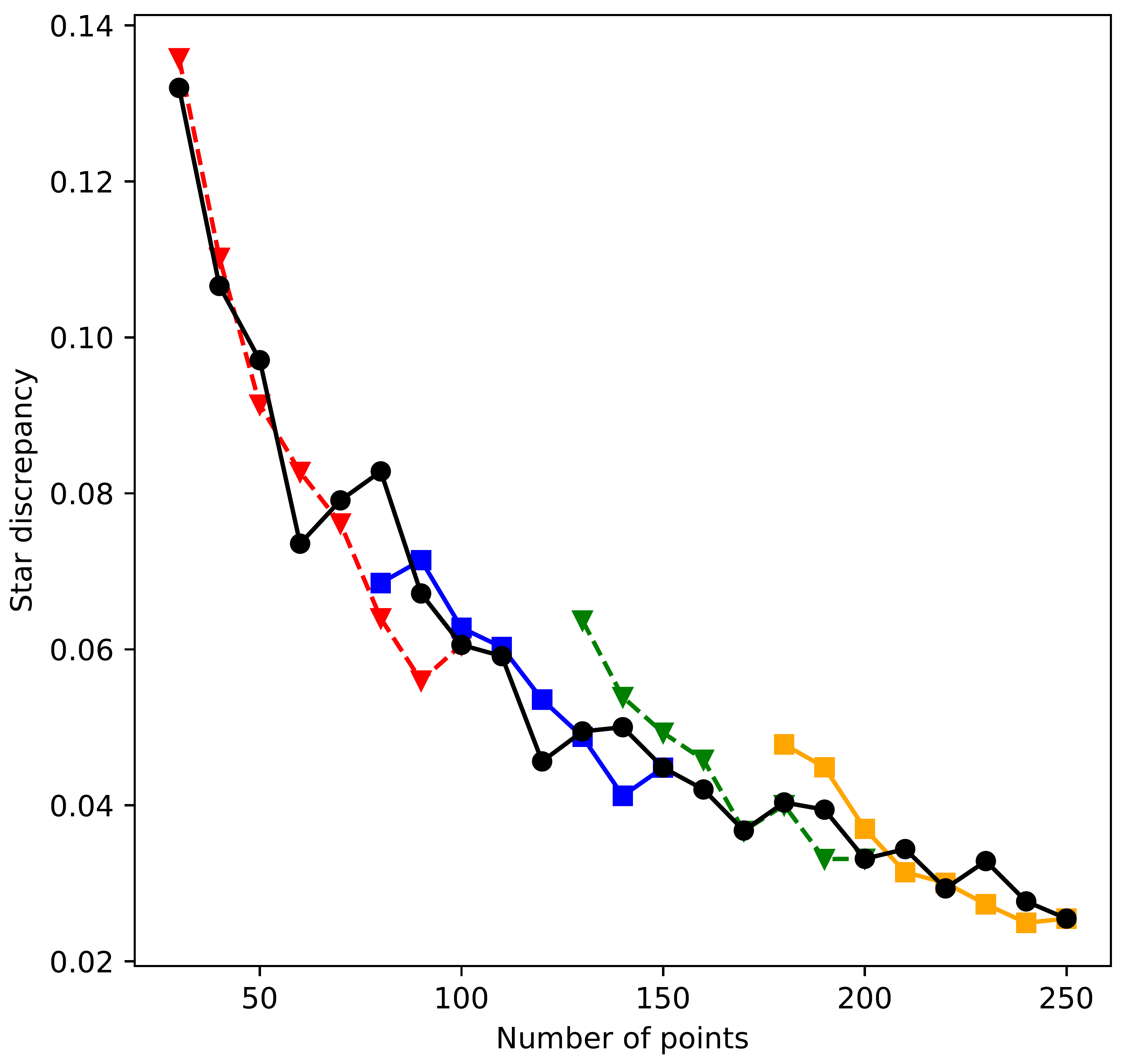}\hfill
\includegraphics[width=0.24\textwidth]{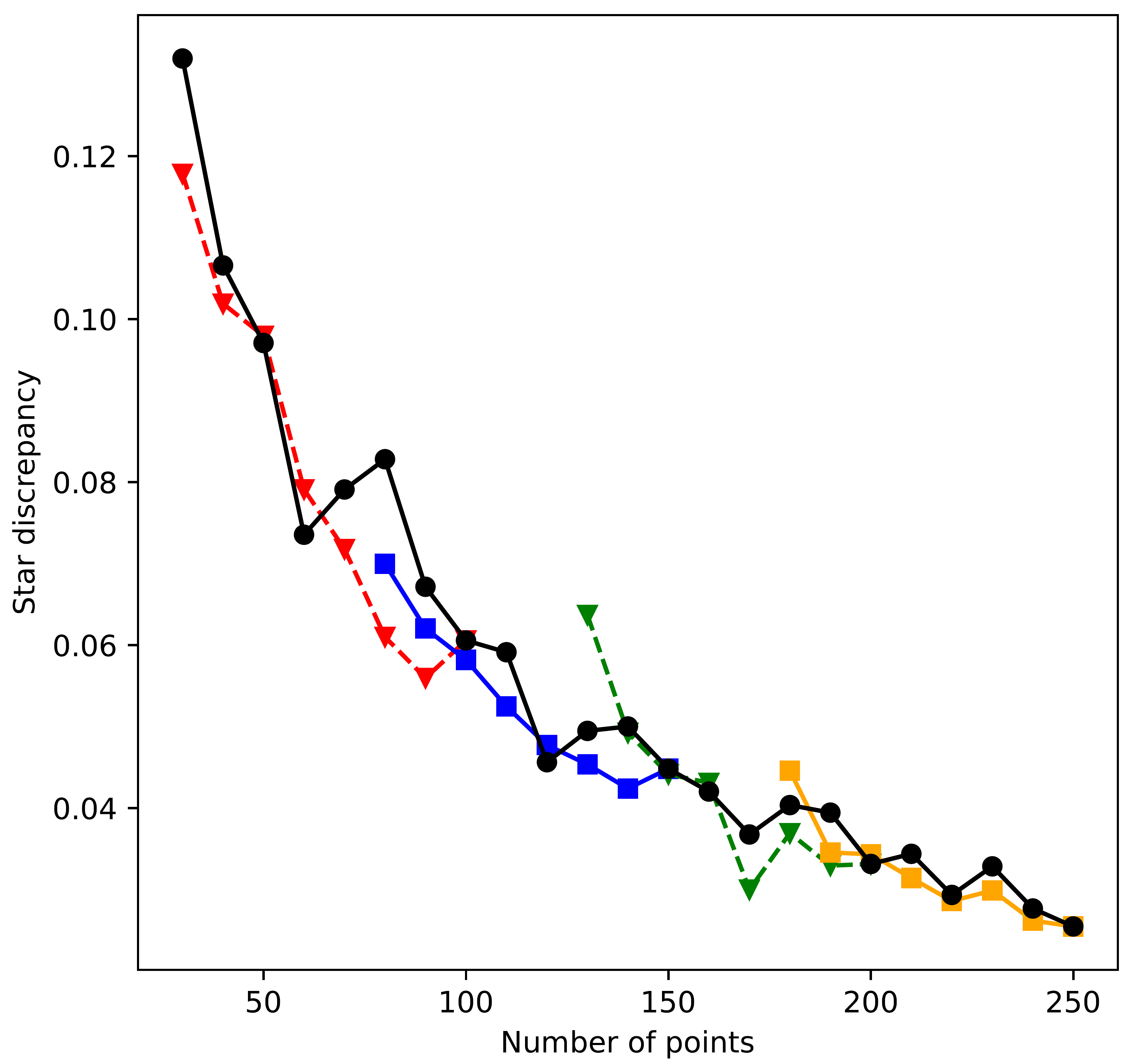}\hfill
\includegraphics[width=0.24\textwidth]{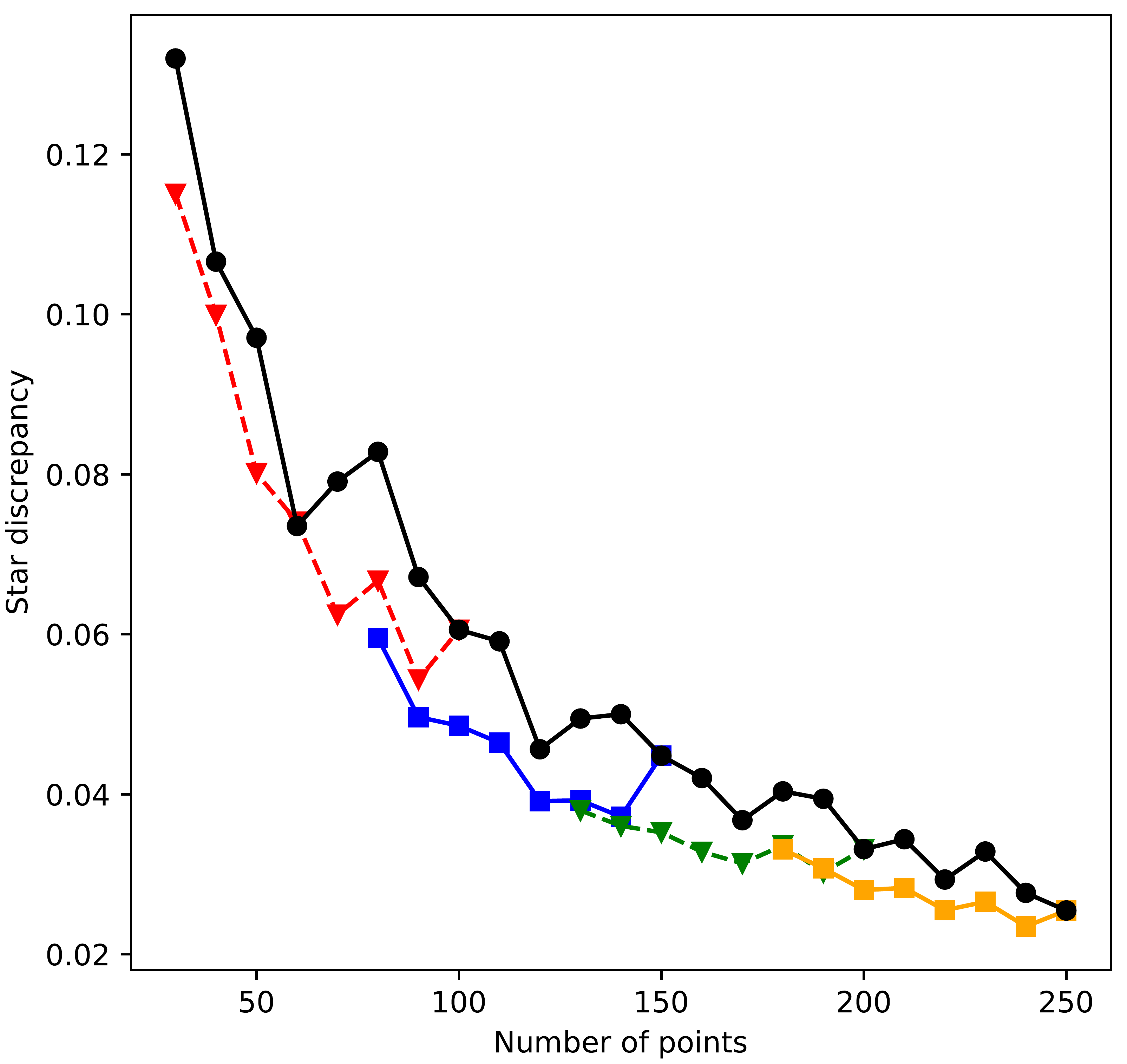}\hfill
\includegraphics[width=0.24\textwidth]{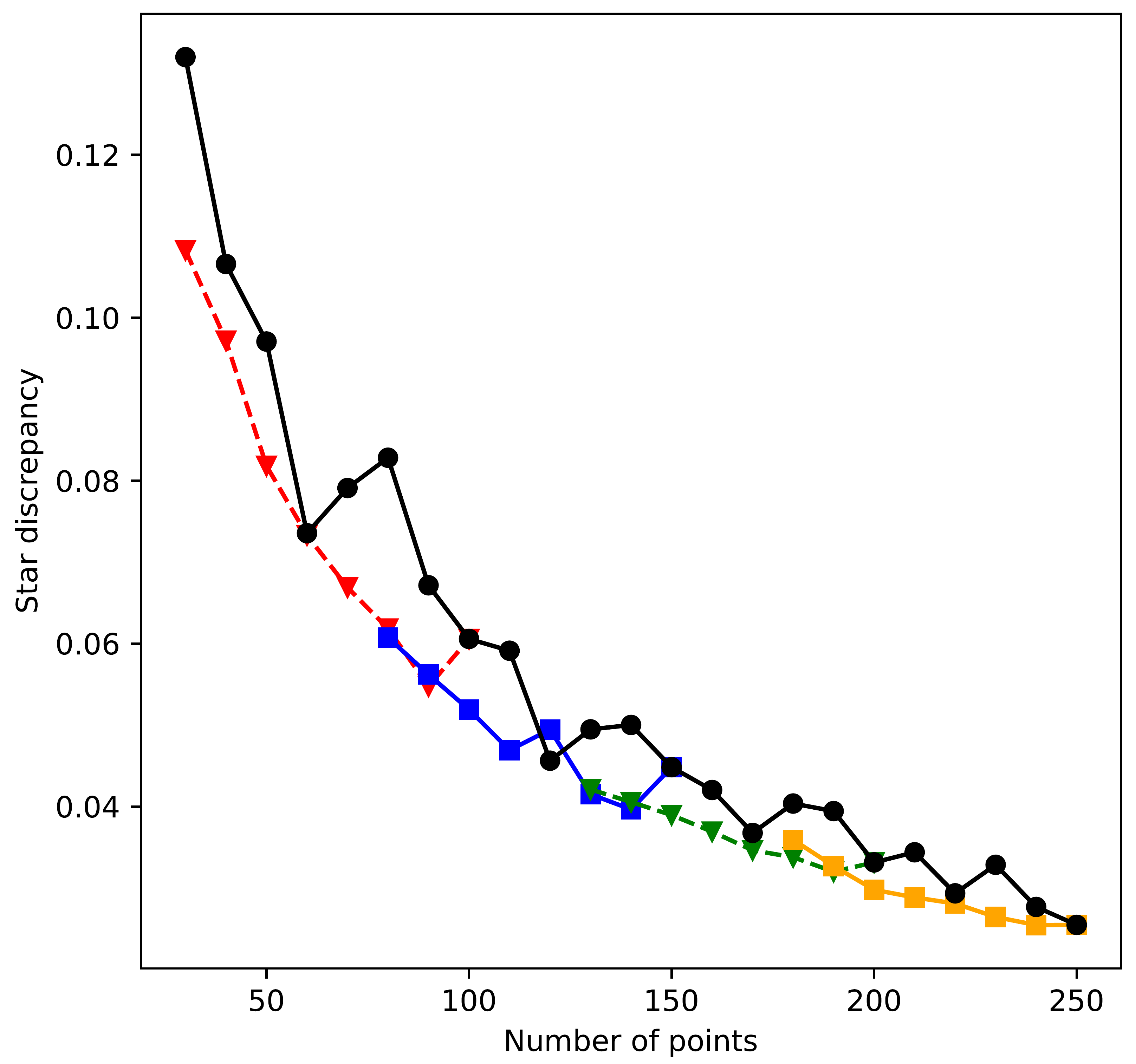}\hfill
\caption{Performance of the different instantiations in dimension 3, from left to right: TA\_BF, TA\_NBF, DEM\_BF and DEM\_NBF. Different colors indicate a change of the initial set size (red for $n=100$, blue for $n=150$, green for $n=200$ and yellow for $n=250$), and the black curve corresponds to the Sobol' sequence (it is the same in all four plots). The plot includes the $k=n$ case for all four different $n$, the rightmost point in this color.}
\label{Dim3}
\end{figure}

\begin{figure}
\centering
\includegraphics[width=0.24\textwidth]{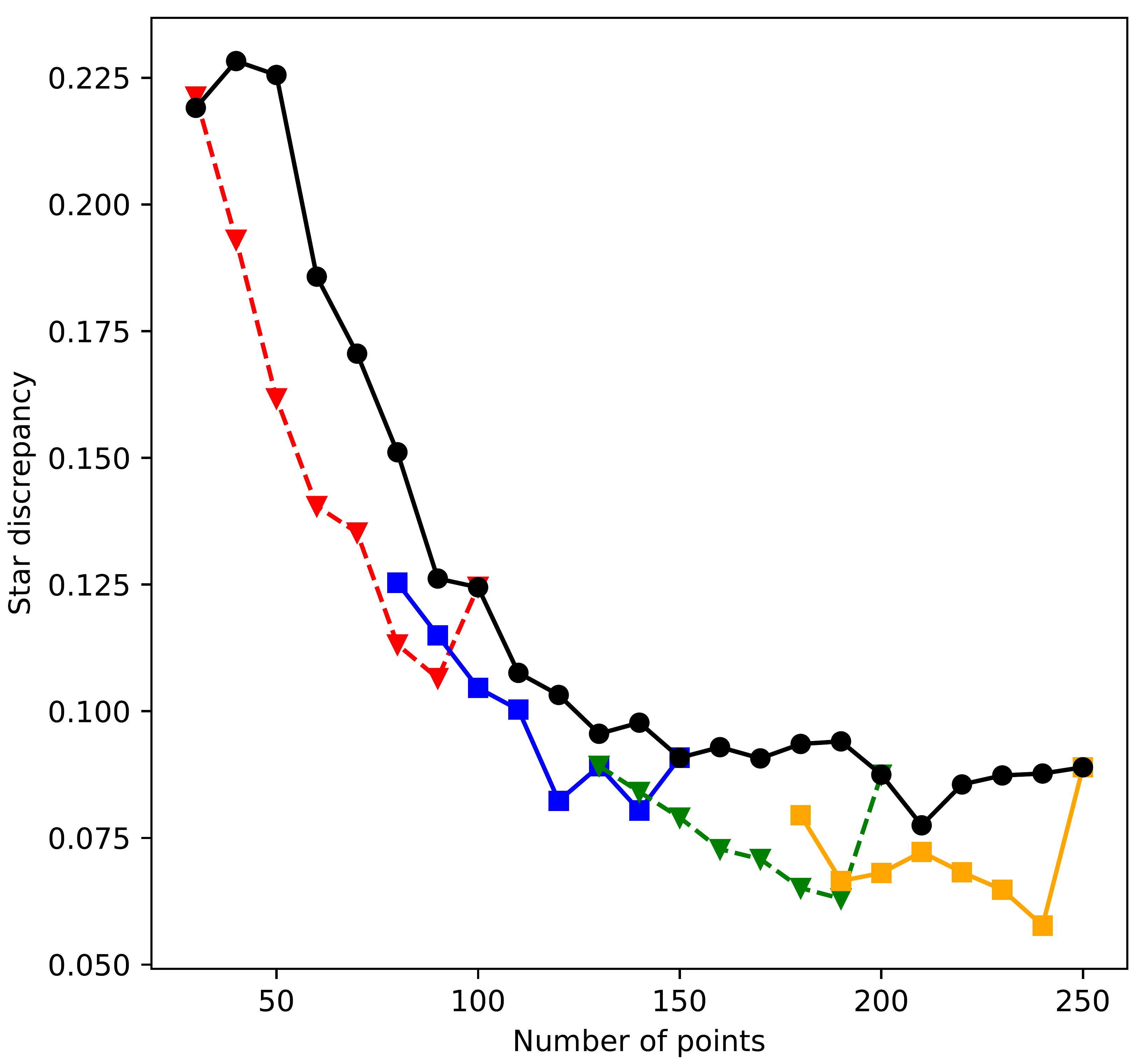}\hfill
\includegraphics[width=0.24\textwidth]{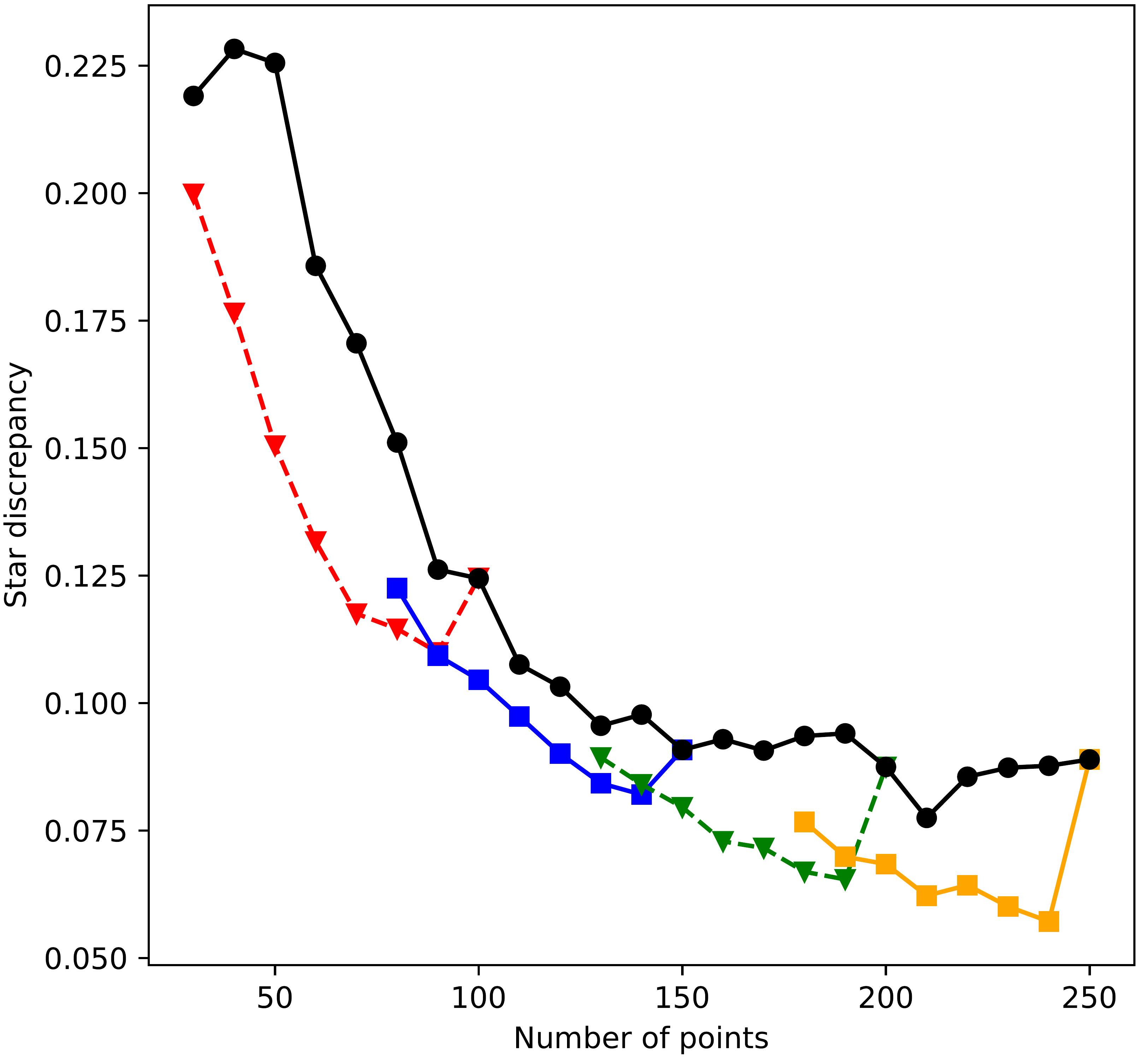}\hfill
\includegraphics[width=0.24\textwidth]{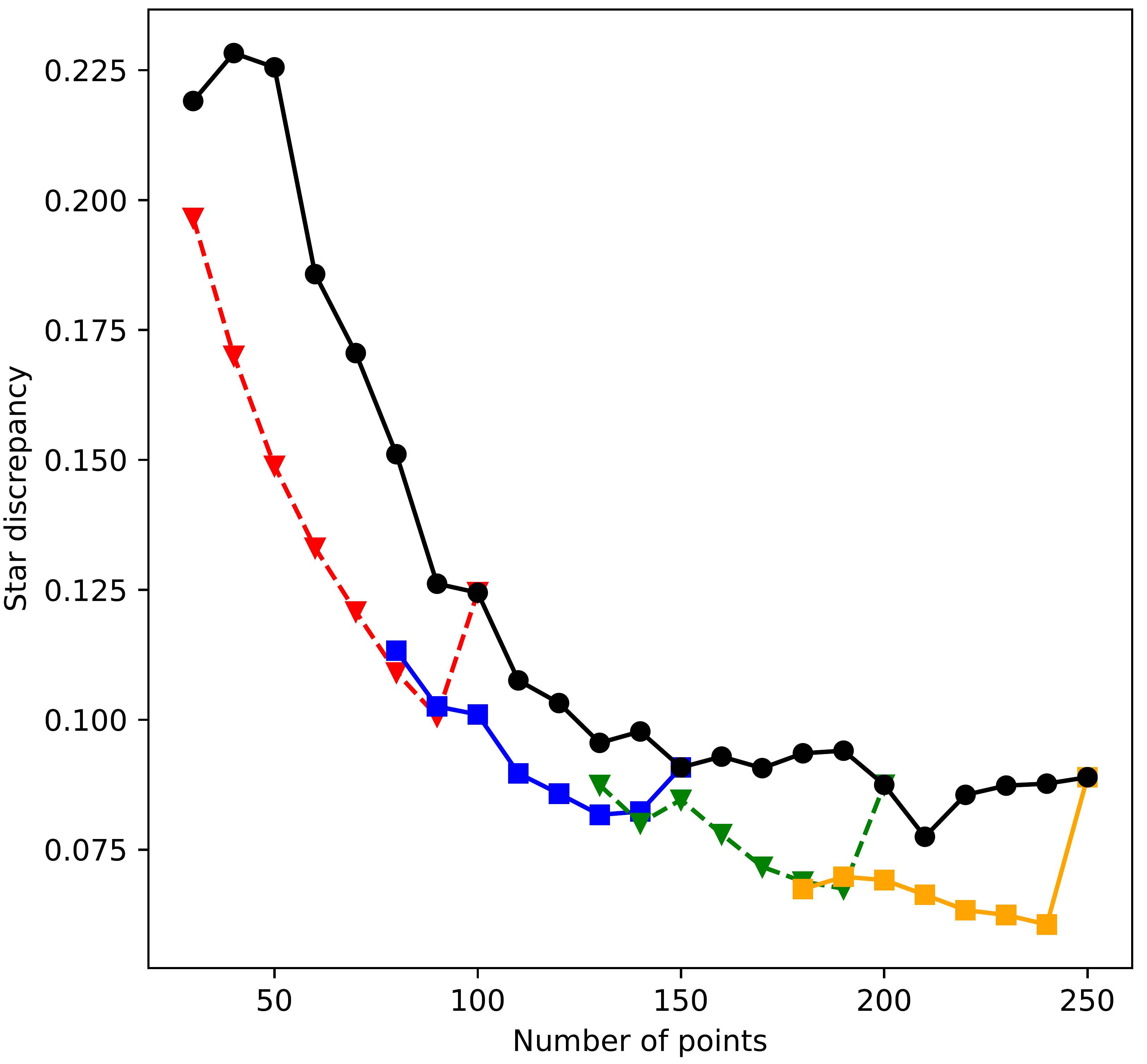}\hfill
\includegraphics[width=0.24\textwidth]{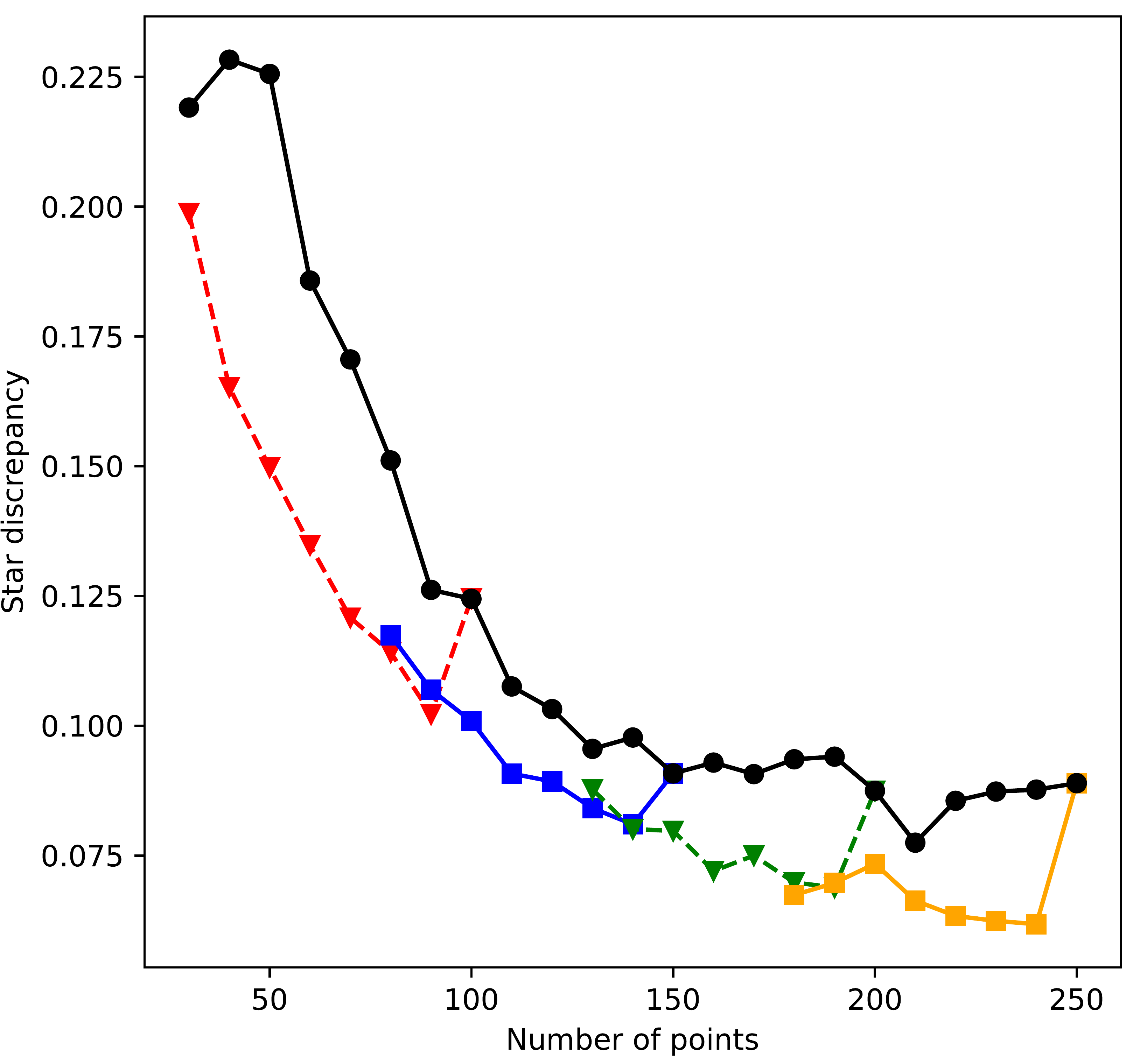}\hfill
\caption{Performance of the different instantiations in dimension 6, from left to right: TA\_BF, TA\_NBF, DEM\_BF and DEM\_NBF. Different colors indicate a change of the initial set size (red for $n=100$, blue for $n=150$, green for $n=200$ and yellow for $n=250$), and the black curve corresponds to the Sobol' sequence (it is the same in all four plots). The plot includes the $k=n$ case for all four different $n$, the rightmost point in each color.}
\label{Dim6}
\end{figure}

\textbf{General tests:} Figures~\ref{Dim3} and~\ref{Dim6} show the performance of our different instantiations in dimensions 3 and 6, respectively. Plots change color to highlight when we are changing the ground set from which these points are selected (i.e., when we increment the previous $n$ by 50). Whenever a heuristic was not able to terminate, we plot the best discrepancy value obtained during the run.

Figure~\ref{Dim3} shows the performance of the different instantiations compared to the Sobol' sequence (black) in dimension 3. DEM\_BF is the best performing version of the heuristic, with  DEM\_NBF outperforming it only for the two smallest instances. For $k \geq 70$, DEM\_BF improves over the Sobol' set of the same size by between 17 and 27\%. DEM\_NBF is also performing better with a 14\% decrease in the discrepancy on average. However, the TA variants are struggling: TA\_NBF improves the Sobol' sequence by only 8\% on average and TA\_BF improves it by 2\%, both being worse than the Sobol' set of similar size for numerous instances. In dimension 3, the DEM algorithm is much faster than the TA heuristic which has a similar runtime regardless of the dimension. This allows the DEM instantiations to run multiple instances within the 1 hour time limit. For example, TA\_BF does not finish even once for $n=200$ and $k=130$, whereas DEM\_BF takes around 41 seconds for a single run. For $n=150$ and $k=80$, DEM\_BF finishes a run in 6 seconds, DEM\_NBF in 0.8 seconds and TA\_NBF in 500 seconds.

Figure~\ref{Dim6} shows that all 4 instantiations have relatively similar performances in dimension 6, all 4 having the best performance on at least one instance. On each instance, the best performing heuristic gives a 10 to 35\% improvement on the discrepancy of the Sobol' sequence of the same size. Even taking the worst performing one, with the exception of 30 points for the TA\_BF heuristic, we have a 7 to 30\% improvement. We note that from $n \geq 120$ and $n-k \geq 30$ onwards, both \_BF instantiations are often (or always for DEM\_BF) unable to finish a single run. We introduced the subset selection problem largely because in higher dimensions a smaller number of samples might not guarantee that low-discrepancy sets would have enough points to reach the asymptotic regime. The tests in dimensions 3 and 6 show that subset selection is more effective in dimension 6, despite the instantiations running fewer tries. For example, DEM\_BF does not finish a single run in dimension 6 but it does 10 separate runs in dimension 3. This seems to confirm our hypothesis that subset selection will perform better in higher dimensions, for which an exponential number of points is required to reach the asymptotic bounds.

\textbf{Fixed $n$ or $k$:} Given these initial results, further tests were done only on the DEM\_BF and TA\_NBF instantiations. In lower dimensions, the DEM algorithm is faster than the TA heuristic. Each heuristic run is fast enough to allow for a brute-force check and the DEM\_BF instantiation gives us the best possible subsets with our method. It also guarantees the correctness of the discrepancy value. For higher dimensions, both the DEM algorithm and the brute-force check become too expensive, TA\_NBF is the only reliably fast instantiation. 

\begin{figure}
\centering
\begin{minipage}{.48\textwidth}
\vspace{0pt}
  \centering
  \includegraphics[width=\linewidth]{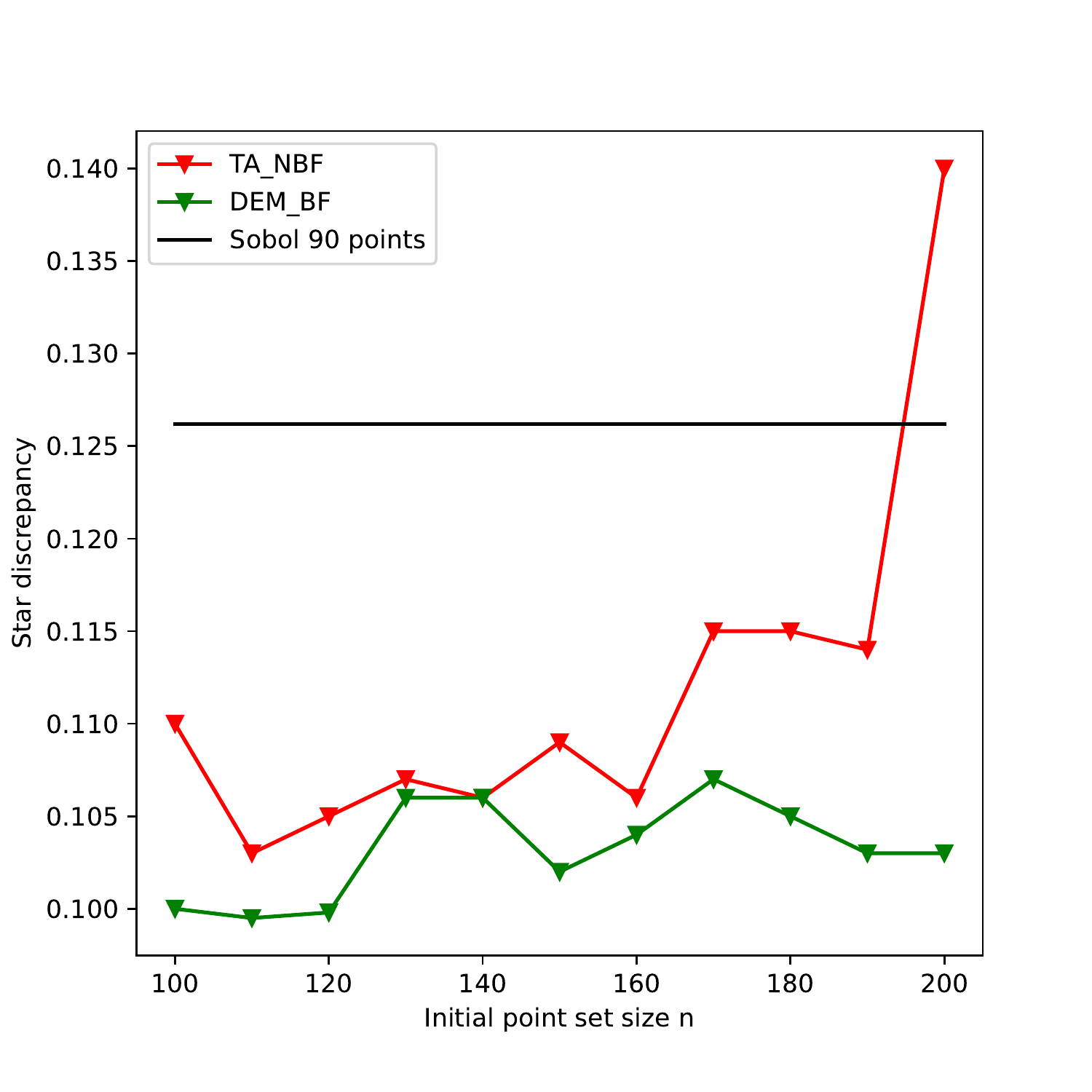}
        \captionof{figure}{Best discrepancy obtained for different values of $n$, $k$ fixed to 90, and $d=6$, with a cutoff time of 1 hour.}
        \label{Fixedk}
\end{minipage}\hfill
\begin{minipage}{.48\textwidth}
  \centering
   \includegraphics[width=\linewidth]{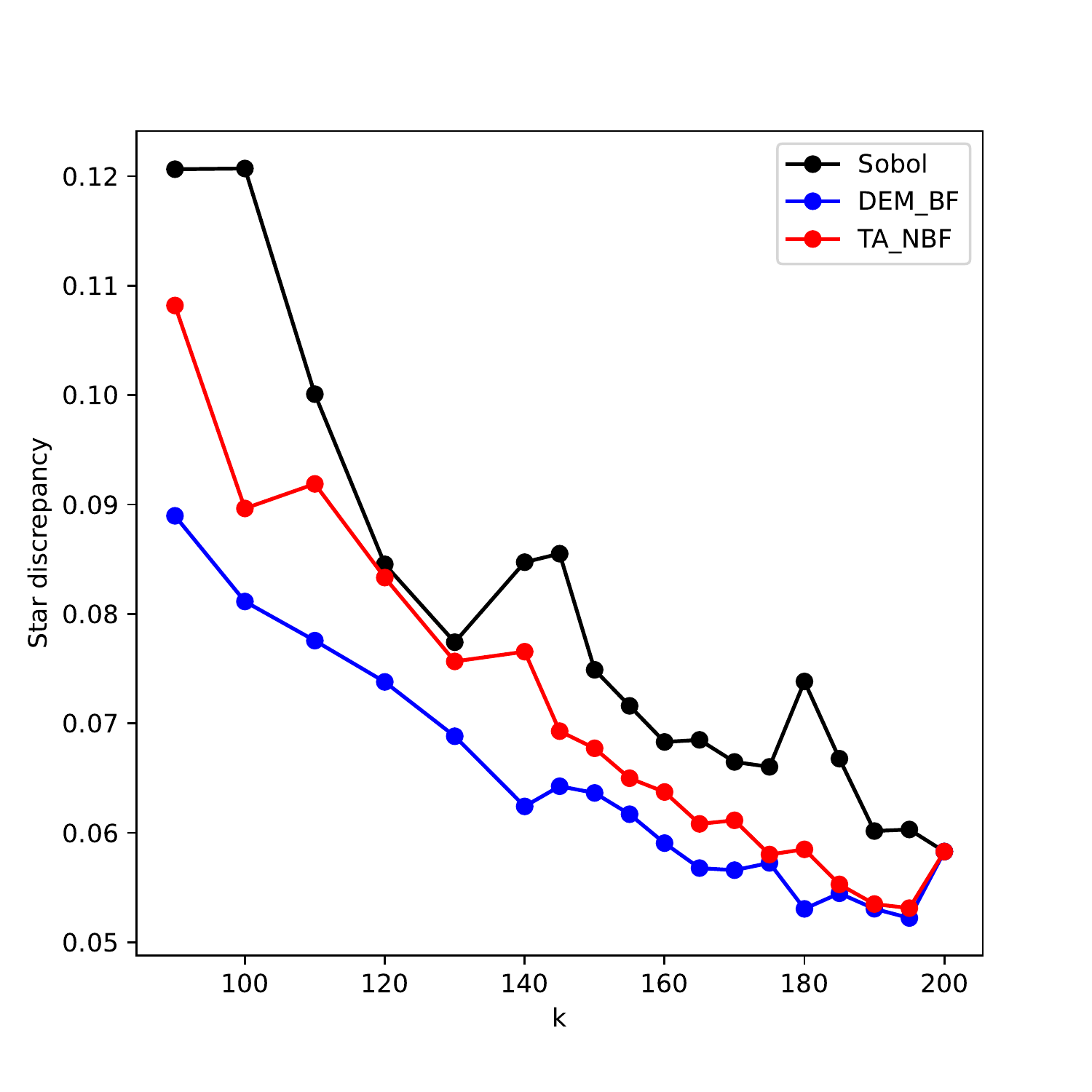}
        \captionof{figure}{Discrepancy obtained for different values of $k$, $n$ fixed to 200 and $d$=5.}
        \label{FixedN}
\end{minipage}
\end{figure}

We now fix the dimension (here $d=6$) and the resulting point set size ($k=90$) to find the best ground set size $n$ to obtain higher quality point sets. In Figure~\ref{Fixedk}, we see that DEM\_BF performs well for all different values of $n$, whereas TA\_NBF works better for $n-k$ close to 20. The first 90 points of the Sobol' sequence have discrepancy 0.126, and only a single instance for TA\_NBF fails to improve this value. The best instances, for DEM\_BF with $n-k \in \{10,20,30\}$, give a 20\% improvement over the Sobol' sequence. These values for $n-k$ are linked to our results on exact methods in~\cite{CDP}, where the greatest discrepancy improvements were also observed for $n-k=20$.

For $n_1>n_2$ and a fixed $k$, the optimal subset selection solution for $n_1$ is better than the one for $n_2$. However, we observe that increasing $n$ is not a good strategy for our heuristic, in particular for TA. We expect this to come from a larger search space which cannot be well explored by the heuristic and from the existence of a large number of sub-optimal local optima.

We then performed a similar experiment with fixed $n$ and  varying $k$ to verify that our results were coherent, this time in dimension 5. The results are shown in Figure~\ref{FixedN}. We first note that the increase in discrepancy when $k$ decreases is expected as we have smaller point sets. The heuristic performs well for all different values of $k$, with both instantiations always outperforming the Sobol' sequence. DEM\_BF improves the Sobol' sequence discrepancy by between 12 and 33\% and TA\_NBF by 2 to 26\%. Once again, TA\_NBF performs much better when the difference between $n$ and $k$ is small whereas DEM\_BF seems to be much more reliable for all values. The discrepancy of the point sets obtained with the heuristics behaves less erratically than the initial Sobol' sequence as the new point sets avoid discrepancy spikes for specific point set sizes. The noticeable improvement obtained by going from 200 to 195 points suggests that removing very few carefully chosen points could also lead to large discrepancy improvements. While we expect this to happen because our heuristics perform better (the search space is much smaller), this could be a promising direction for cheaper methods of improving low-discrepancy point sets. 

\begin{figure}
\centering
\begin{minipage}{.48\textwidth}
\vspace{0pt}
  \centering
  \includegraphics[width=\linewidth]{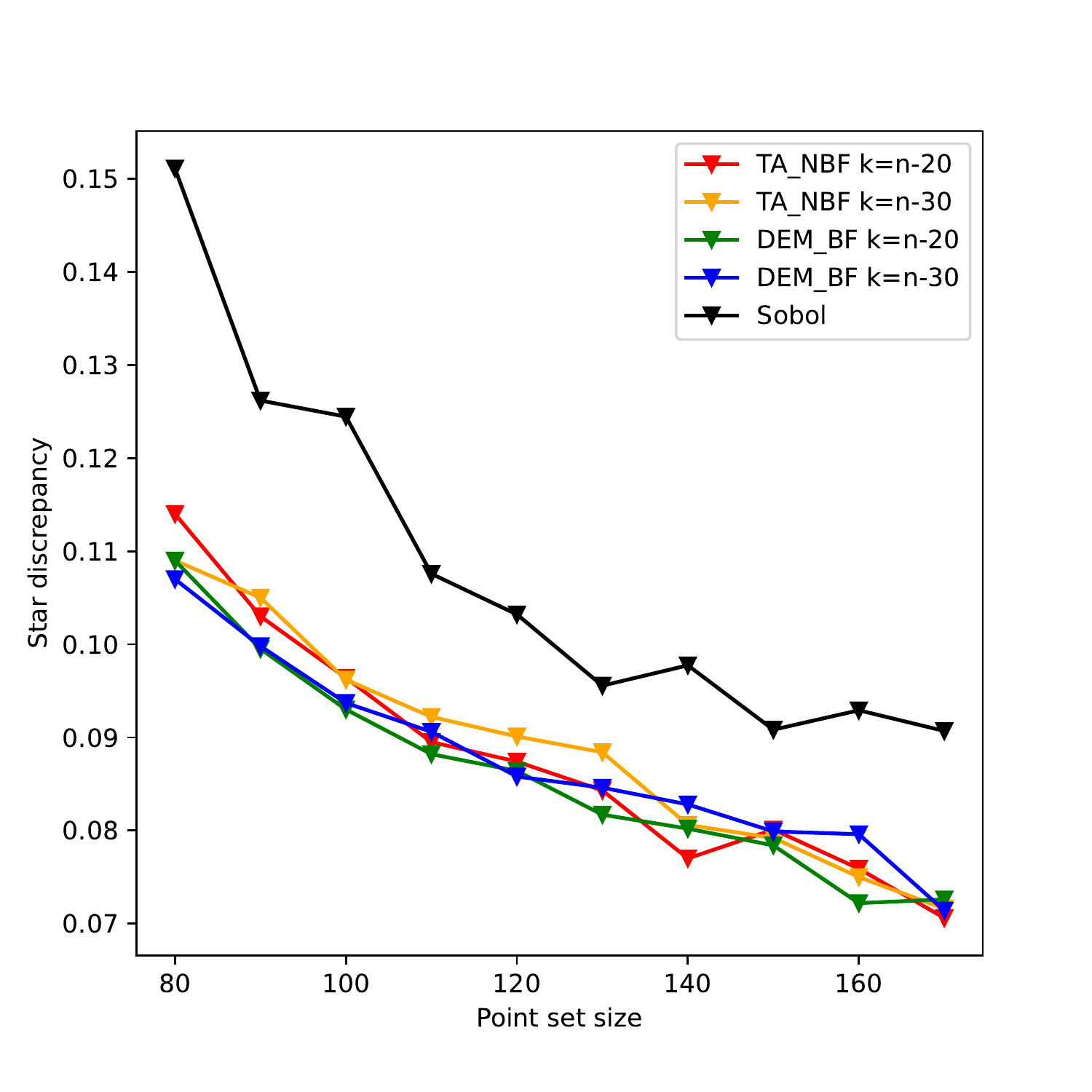}
        \captionof{figure}{Best discrepancy values obtained with our heuristic for $d=6$ and k=$n-20$ or $k=n-30$}
        \label{Opti6}
\end{minipage}\hfill
\begin{minipage}{.49\textwidth}
  \centering
   \includegraphics[width=\textwidth]{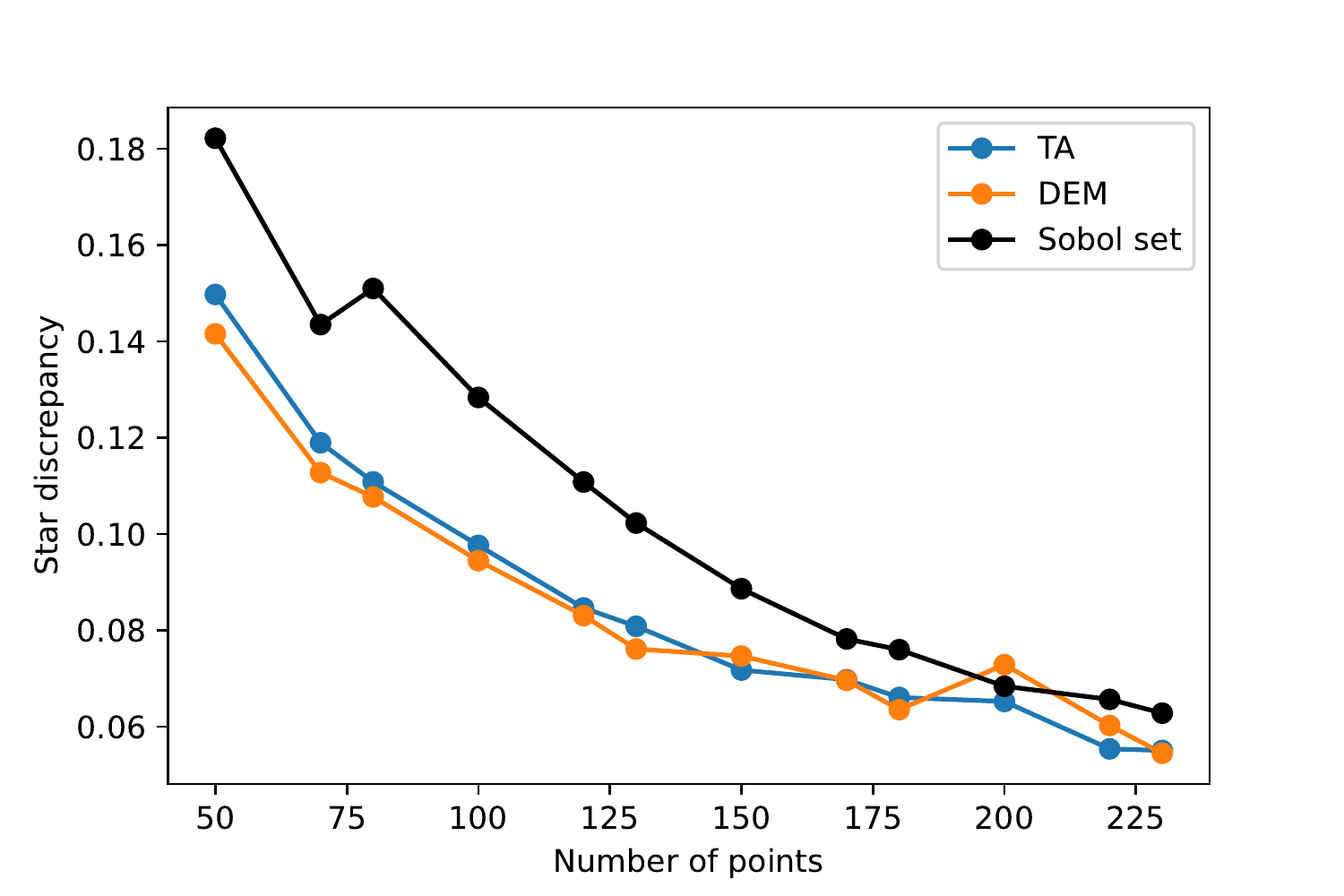}
    \captionof{figure}{Performance of the two subset selection instantiations on Sobol' sets in dimension 6. Subset selection was done for $k=n-20$, $n-30$ and $n-50$ and $n\in \{50,100,150,200,250\}$.}
    \label{fig:set}
\end{minipage}
\end{figure}

\textbf{Best results:} Figure~\ref{Opti6} gives the best values obtained by TA\_NBF and DEM\_BF for $k=n-20$ and $k=n-30$, which should be the best conditions for our heuristic given the previous results. We notice that there is very little difference between both algorithms, with $k=n-20$ or $k=n-30$. In all cases, our heuristic is clearly outperforming the Sobol' sequence, the discrepancy value for 170 points of Sobol' is reached at 120 or 130 points for all our point sets. Our heuristic improves the Sobol' sequence's discrepancy by 8 to 30\% depending on the instantiation choice. For each choice of $k$ and $n$, the worst instantation improves by between 8 and 25\% the discrepancy, whereas the best performing one improves by 15 to 30\%. While the plots here show results in dimension 6, our experiments show that our heuristic performs well for all dimensions for which we can compute the discrepancy. We note that results become poorer for much larger $n$: if $n=500$ and $k=480$, the benefit of using subset selection becomes quite small.~\ref{App} gives a greater set of values obtained during our experiments.

\textbf{Comparison with low-discrepancy point sets}: We note that subset selection is not limited to low-discrepancy sequences but can also be used with a low-discrepancy set. One can obtain an $n$-point set in dimension $d+1$ from a low-discrepancy sequence in dimension $d$ by taking the first $n$ points and adding $i/n$ as the $d+1$-th coordinate of the $i$-th point. Figure~\ref{fig:set} shows the results obtained by using subset selection on the obtained sets in dimension 6, starting from the Sobol' sequence. The results are similar as those in the sequence case: $k=n-20$ and $k=n-30$ give the best results, while the improvement in the discrepancy value is up to 33\%. Finally, the discrepancy values between the set version of Sobol' obtained with the $d$-dimensional sequence and the $d+1$-dimensional sequence are quite similar. This suggests that subset selection on \emph{sequences} provides point sets better than low-discrepancy \emph{sets}.

\textbf{Comparison with random subset selection:} Finally, we note that selecting the best subset from a large number of random subsets does not work well. This had been tested extensively when comparing with the exact case in~\cite{CDP}, and remains true here. We only provide some simple examples. For $n=100$ and $k=80$ in dimensions 4 and 5, 100\,000 random subsets give us a best discrepancy of respectively 0.081502 and 0.099460, roughly 10\% worse than the DEM with brute force instantiation. This becomes even worse when $n-k$ increases and the number of possible subsets increases: for $n=100$ and $k=60$ in dimension 4 the best random subset has discrepancy 0.105830, against 0.087650 with subset selection. While these values do not seem too bad, the main cost of our algorithm is calculating discrepancies. In our experiments, the values were obtained with between 10\,000 and 20\,000 discrepancy evaluations in the brute force cases and 1\,000 and 2\,500 evaluations without brute force: far less than the 100\,000 required with random subsets to obtain half or a third of the improvement.

\subsection{Improvements for the Inverse Star Discrepancy}\label{sec:inverse}

Obtaining point sets with better star discrepancy naturally leads to improvements for the inverse star discrepancy problem: given a discrepancy target $\varepsilon$, what is the minimal $n$ such that there exists a point set $P$ such that $|P|=n$ and $d_{\infty}^{*}(P) \leq \varepsilon$? Table~\ref{inverse1} shows the improvement in inverse star discrepancy by using the subset selection approach. For applications where the evaluation of each point can take a whole day of calculations, the 12 to 46\%  gain is substantial. These values were obtained by only taking results from our previous experiments (either in figures or in the tables in~\ref{App}). We did not try to find the smallest values to reach these discrepancy targets. It is likely that our results could be further improved by a more targeted search. Adapting parameters to the desired instance, increasing the number of runs or removing the 1-hour cutoff are possible options to obtain better results.

\begin{center}

\begin{table}[]
    \centering
    \caption{Number of points necessary to reach target discrepancies for subset selection and Sobol' in dimensions 4 and 5}
    \begin{tabular}{|c c c c|}
    \hline
        Dimension & Target discrepancy & Sobol' $n$ & subset selection $n$  \\
        \hline
         $d=4$ & 0.30 & 15 & 10 \\
         & 0.25 & 17 & 15 \\
         & 0.20 & 28 & 20\\
         & 0.15 & 45 & 30\\
         & 0.10 & 89 & 50\\
         & 0.05 & 201 & 170\\
         \hline
         $d=5$ & 0.30 & 17 & 10\\
         & 0.25 & 26 & 20\\
         & 0.20 & 38 & 25\\
         & 0.15 & 52 & 40\\
         & 0.10 & 112 & 70\\
         & 0.05 & 255 & 210\\
         \hline
    \end{tabular}
    \label{inverse1}
\end{table}
\end{center}

\begin{figure}
\begin{minipage}[t]{.48\textwidth}
\centering
\vspace{0pt}
  \includegraphics[width=\linewidth]{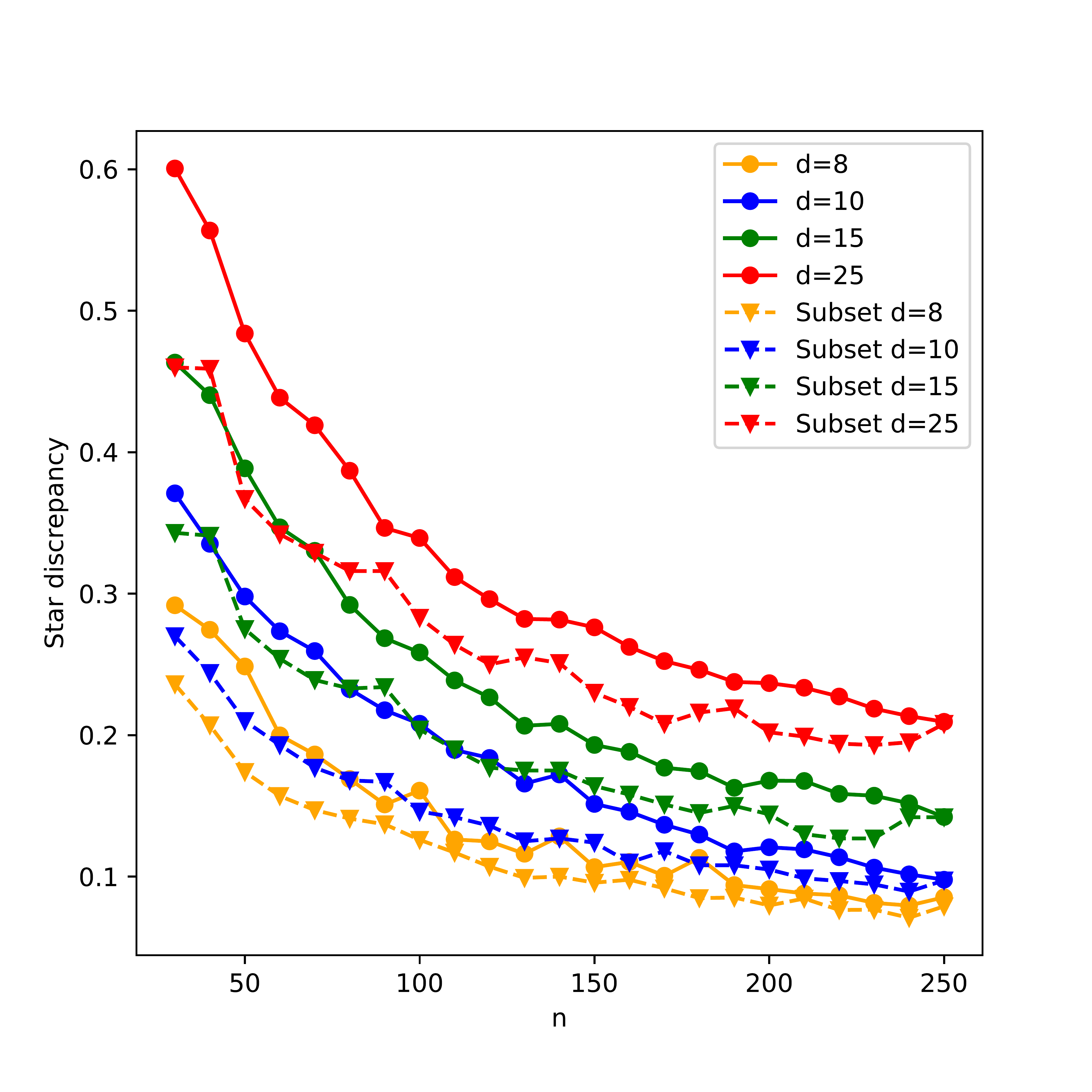}
  \captionof{figure}{Discrepancy values obtained for the Sobol' sequence in different dimensions, compared with the values obtained by subset selection (dashed lines, using the TA heuristic).}
  \label{Discs}
\end{minipage}\hfill
\begin{minipage}[t]{.48\textwidth}
\centering
\vspace{0pt}
  \includegraphics[width=\linewidth]{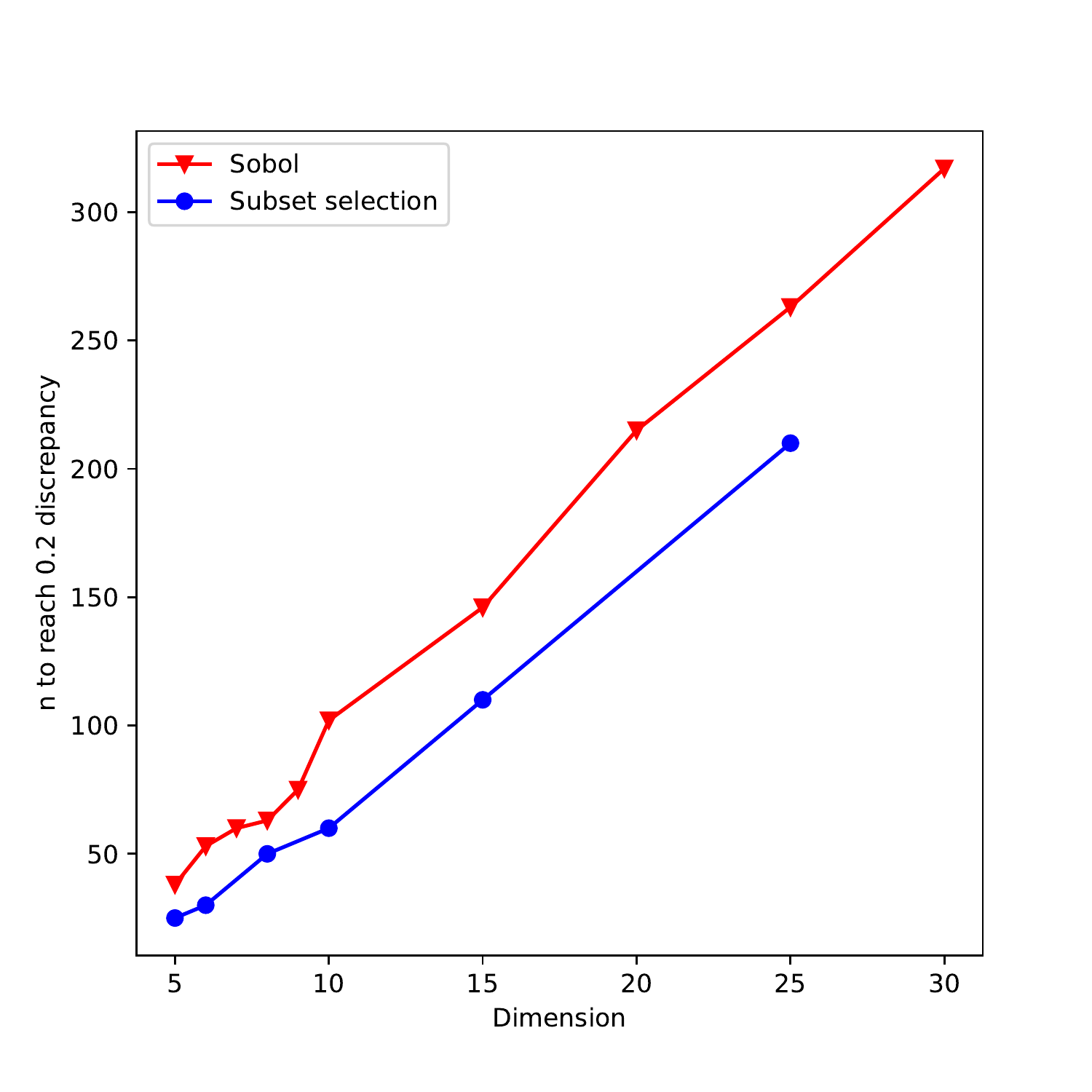}
  \captionof{figure}{Number of points needed to obtain a discrepancy of 0.2 in different dimensions.}
  \label{Epsi2}
\end{minipage}

\end{figure}

We note that the star discrepancy of some known sequences may have been theoretically overestimated, or most likely simply never calculated. In~\cite{NW}, Open Problem 42 lists three open questions with targets for the inverse star discrepancy, as well as a conjecture that $n=10d$ would be a sufficient number of points to reach $d_{\infty}^{*}=0.25$. The open questions had been solved by Hinrichs~\cite{Hinrichs} for the first, and later by Doerr and de Rainville~\cite{Rainville} for all three, each time by building a new point set. Our experiments on the Sobol' sequence show that $n=7d$ points are sufficient at least for lower dimensions (smaller than 10). Figure~\ref{Discs} shows how the discrepancy of the Sobol' sequence evolves for specific dimensions, as well as a comparison with our subset selection sets. 

All three open problems seem to be solved by taking a few hundred points rather than the thousands suggested, confirming de Rainville and Doerr's results without requiring a new set construction. For example, in dimension 15, 146 points have a discrepancy of 0.198, in dimension 30 320 points have a discrepancy of 0.193 and in dimension 25 1205 points have a discrepancy of 0.0996 which can then be lifted to a 50-dimensional point set with discrepancy smaller than 0.2 using Hinrichs' lifting procedure~\cite{Hinrichs}. We acknowledge that these discrepancy values may not be exact as we used the TA heuristic to compute them. However, there is such a large margin both in discrepancy value and number of points that we believe the Sobol' sequence to solve the three problems posed in~\cite{NW}. 

Figure~\ref{Epsi2} shows the number of points of the Sobol' sequence that are needed to reach discrepancy less than or equal to 0.2 depending on the dimension. This number of points should not be seen as an exact value, but as an upper-bound (with the TA imprecision caveat). Indeed, we found this via binary search to avoid having to compute discrepancy values for all possible $n$. However, since the star discrepancy of the Sobol' sequence is not monotonous in~$n$ (see Figure~\ref{Opti6}), it is possible we have missed better point sets. Nevertheless, we observe a linear relation between the dimension and the number of points, close to $n=10d$. This reinforces our impression that even the $n=10d$ conjecture from~\cite{NW} is overestimating the number of points necessary to reach discrepancy $ \leq 0.25$.

\subsection{Comparison with the Energy Functional} \label{CompaSte}

In~\cite{StefEnerg}, Steinerberger introduced the following functional for a point set $X=(x_i)_{i \in \{1,\ldots,N\}}$

\begin{equation}
    E[X]:= \sum_{\substack{1 \leq m,n \leq N \\ m \neq n}} \prod_{k=1}^{d} (1-\log(2 \sin(|x_{m,k}-x_{n,k}| \pi))).
\end{equation}
This expression was derived from the Erd{\H{o}}s-Koksma-Turán inequality~\cite{EKT1,EKT2,EKT3}, modified to allow the use of gradient descent for optimization. Starting with a given point set of any kind, he applied standard gradient descent until convergence to obtain a new point set which should be better distributed and hopefully have lower discrepancy. He provided a number of examples in dimension 2, while underlining that specific point sets could not be improved by his functional. We ran more extensive experiments in higher dimensions, especially in a setting where $n$ is not necessarily far larger than $d$. We implemented the functional in C. 

\begin{figure}
    \centering 
    \includegraphics[trim={0cm 17.0cm 4.0cm 1cm},clip,width=0.9\linewidth]{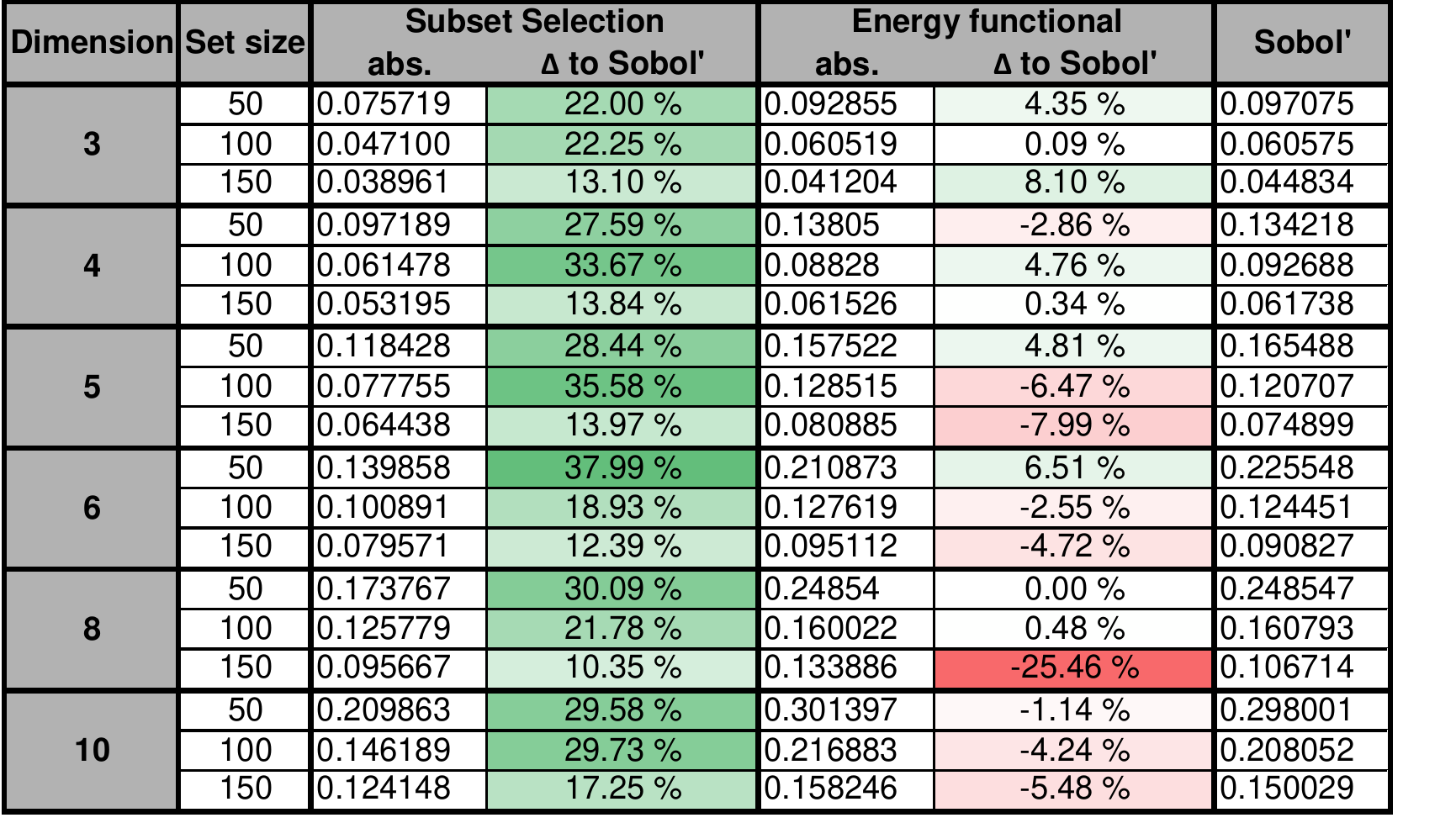}
    \caption{Comparison of the energy functional and subset selection. The functional is applied to the Sobol' set of the same size, the subset selection results are taken from our general experiments with $n-k=50$ for DEM\_BF or TA\_NBF. Also showing percentage improvement of subset selection and the energy functional compared to the Sobol' sequence.
    }
    \label{fig:compare}
\end{figure}

Figure~\ref{fig:compare} shows our results and compares the obtained point sets with the results obtained via our subset selection approach. While the energy functional manages to improve in most cases the discrepancy of the input point set, it is much less effective than subset selection. In particular, it is sometimes unable to improve the Sobol' sequence, for example for $d=5$ and $k=100$. Subset selection is therefore more effective than the functional at creating a new point set with lower discrepancy.

\begin{figure}
    \centering 
    \includegraphics[trim={0cm 18.5cm 4cm 1cm},clip,width=0.9\linewidth]{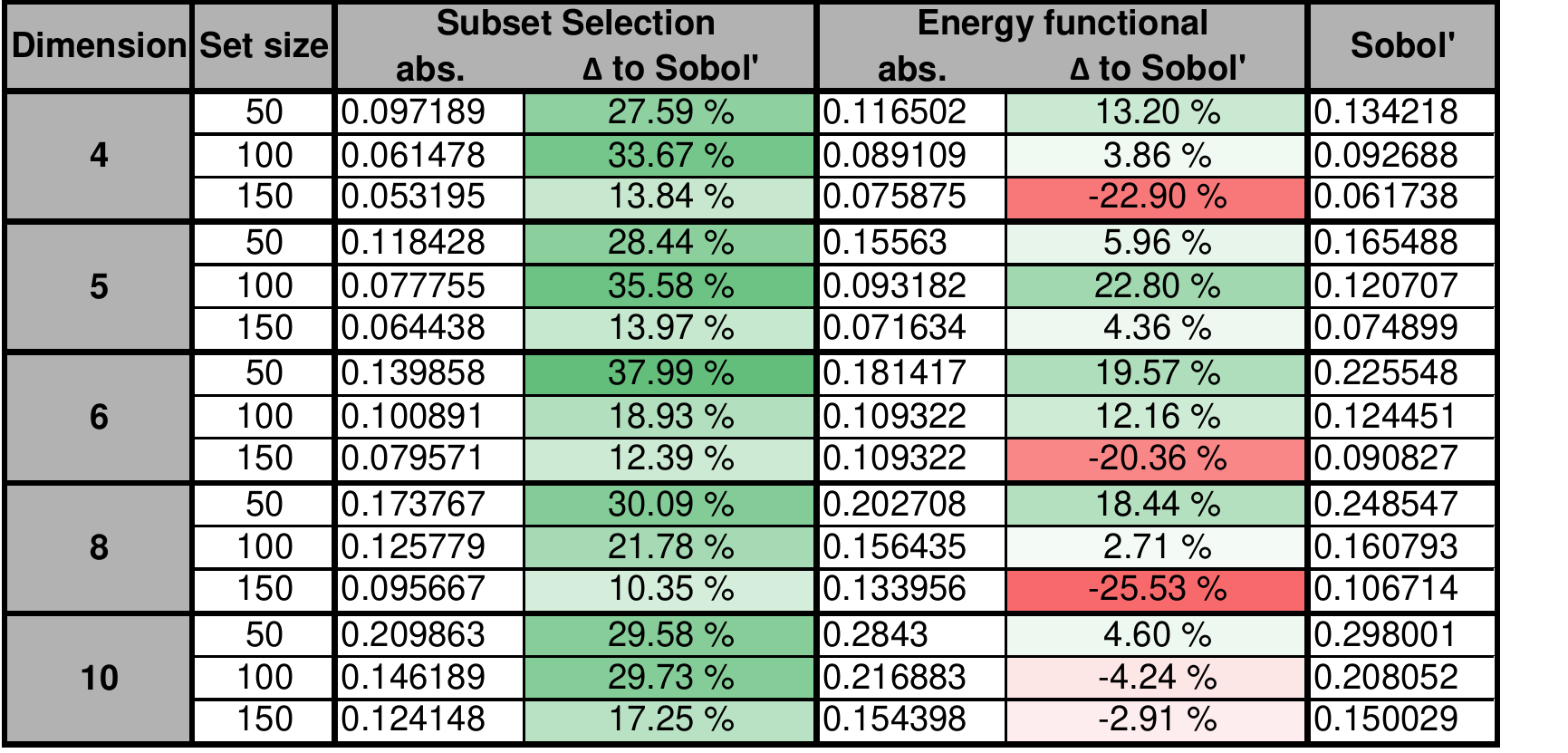}
    \caption{Applying the energy functional to our subset selection point sets. These point sets were obtained with $k=n-50$, see~\ref{App}. Also showing percentage improvement of subset selection and subset selection+energy functional compared to the Sobol' sequence of the same size.
    }
    \label{fig:SeqFunc}
\end{figure}

Furthermore, the functional cannot be applied to our new point sets to obtain better point sets. Applying the energy functional optimization to our own low-discrepancy point sets makes them noticeably worse, removing a large part of the initial gain of subset selection. Figure~\ref{fig:SeqFunc} gives the discrepancies of point sets obtained by DEM\_BF or TA\_NBF ($n$ is the nearest multiple of 50 in each line), to which we apply the energy functional to obtain the point sets for the third column. The Sobol' point sets are added as a comparison point in the last column. We observe that the functional makes the point sets noticeably worse, sometimes even worse than the Sobol' point set of corresponding size. This also shows that the energy functional cannot be used as a surrogate for the discrepancy, the point sets obtained with the energy functional approach have much lower energy than those found by subset selection.

\begin{figure}[t]
    \centering
    \includegraphics[width=.32\textwidth]{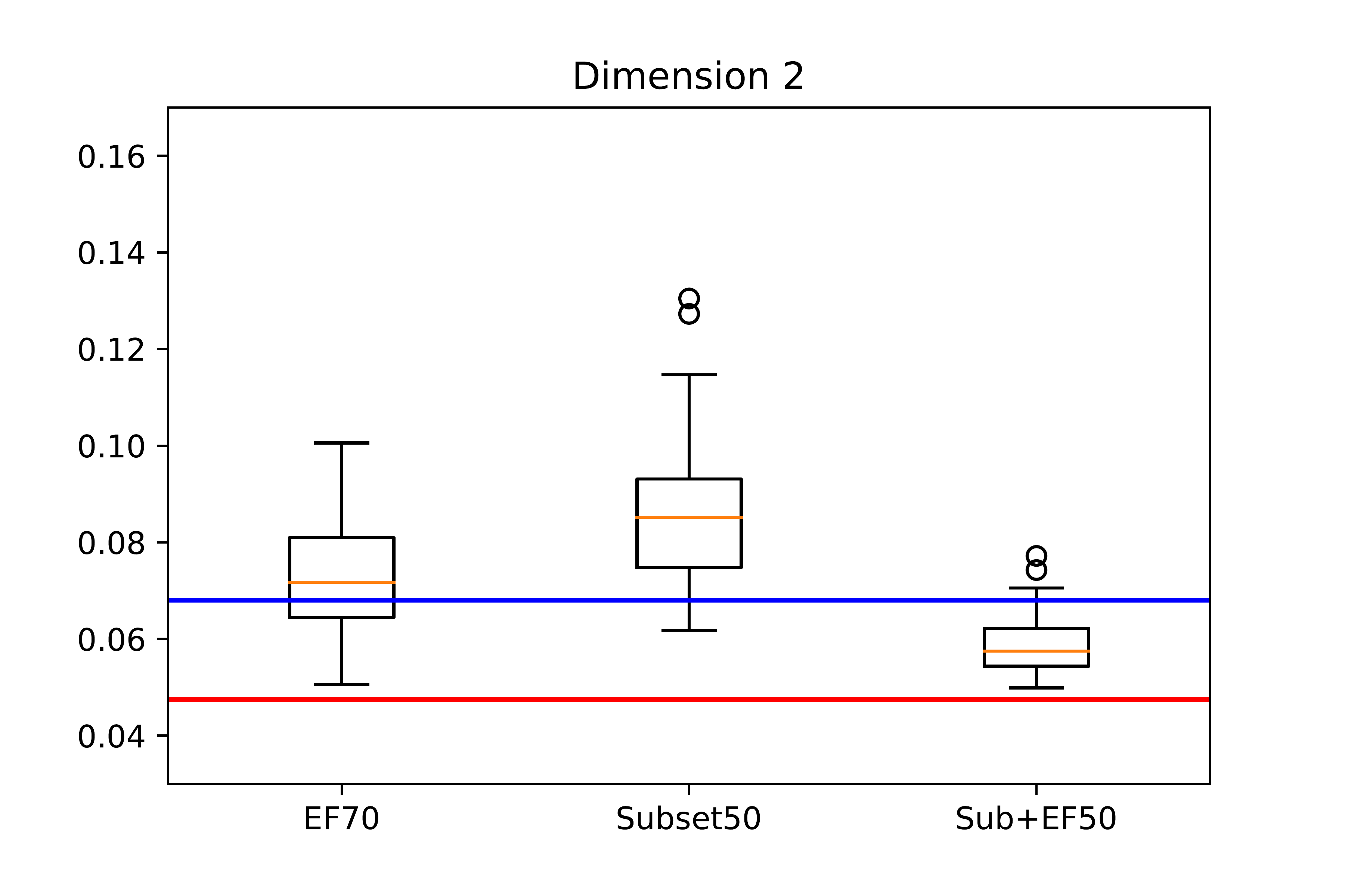}\hfill
    \includegraphics[width=.32\textwidth]{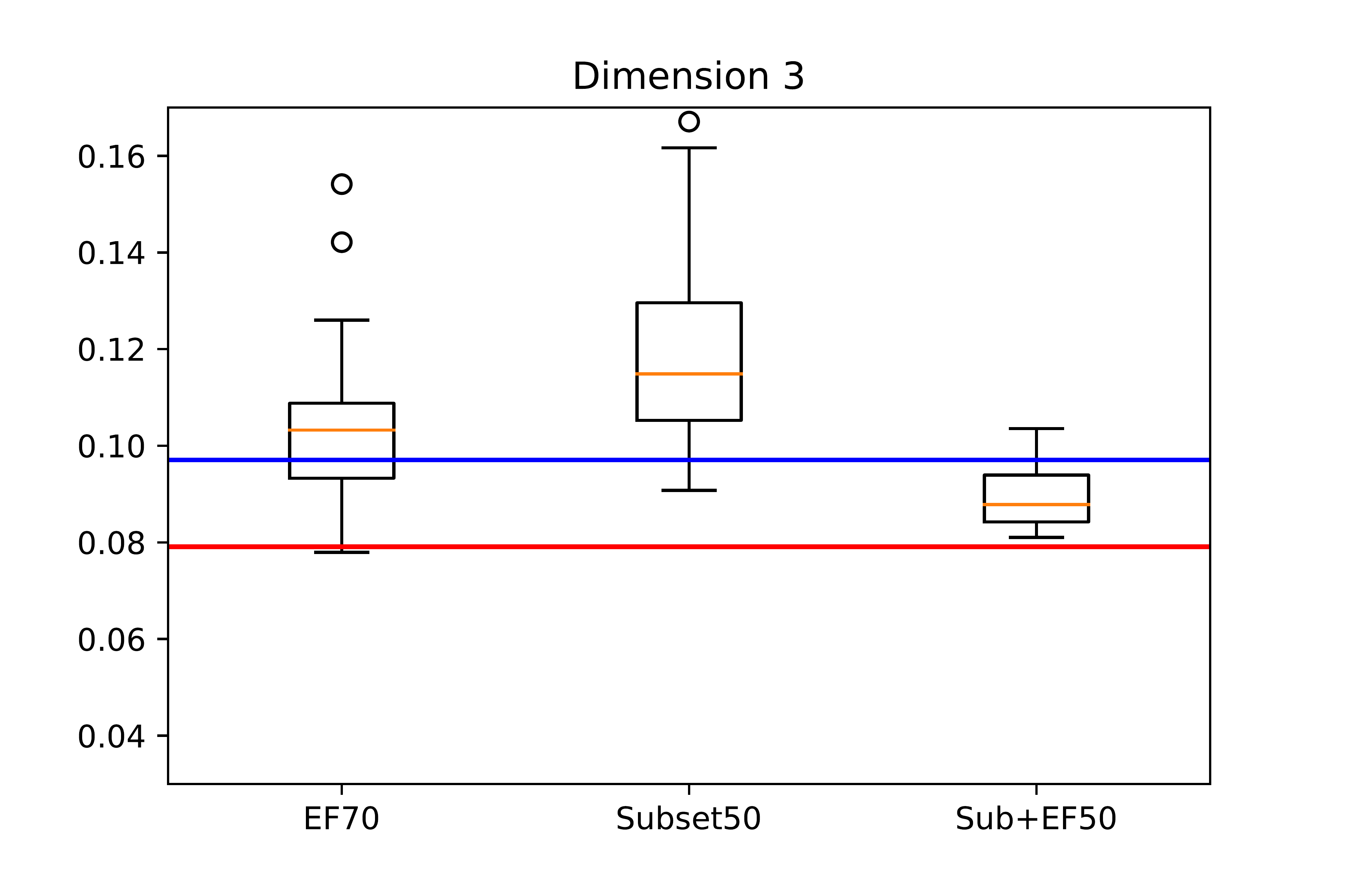}\hfill
    \includegraphics[width=.32\textwidth]{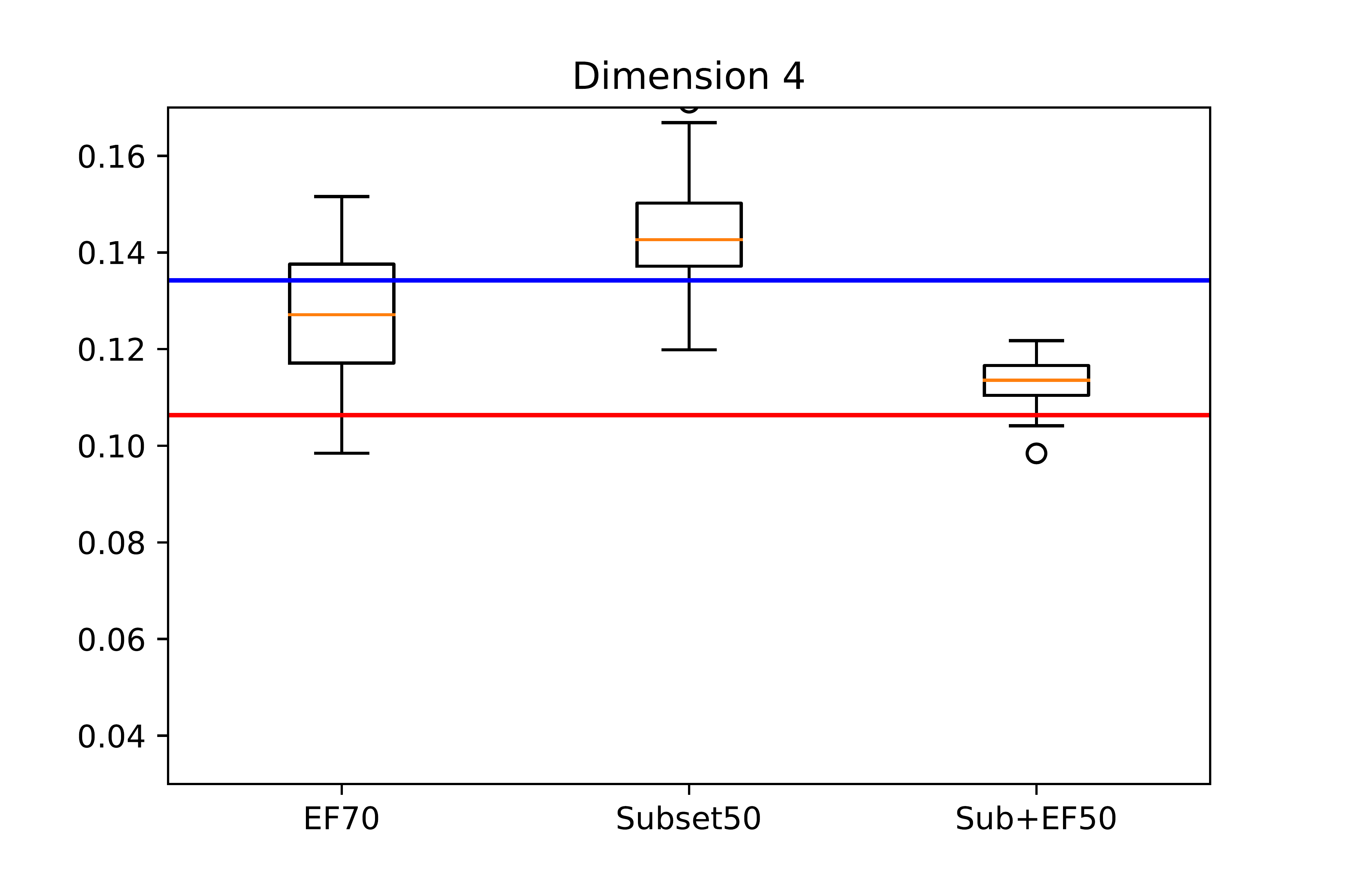}\hfill \smallskip
    \includegraphics[width=.32\textwidth]{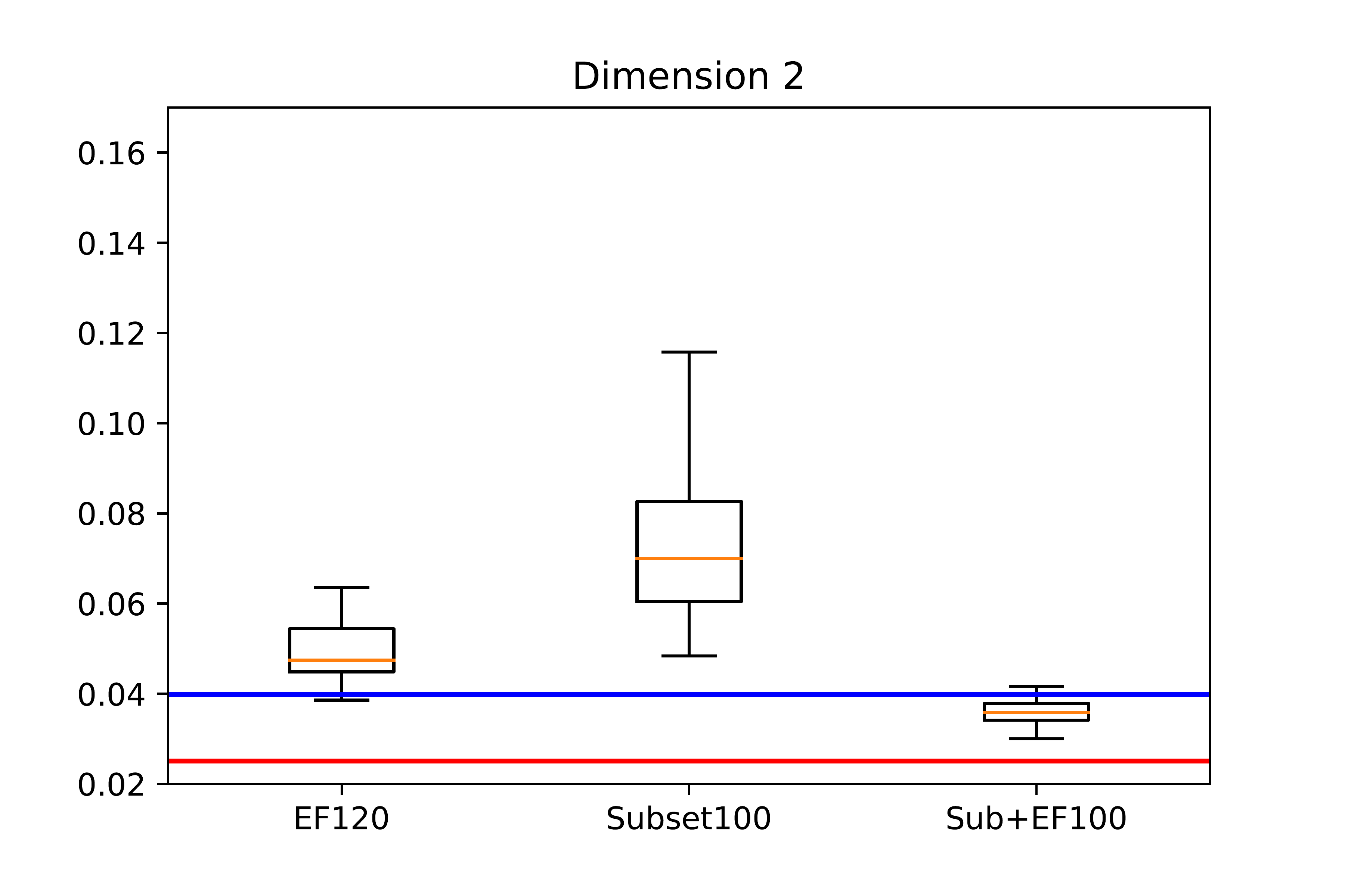}\hfill
    \includegraphics[width=.32\textwidth]{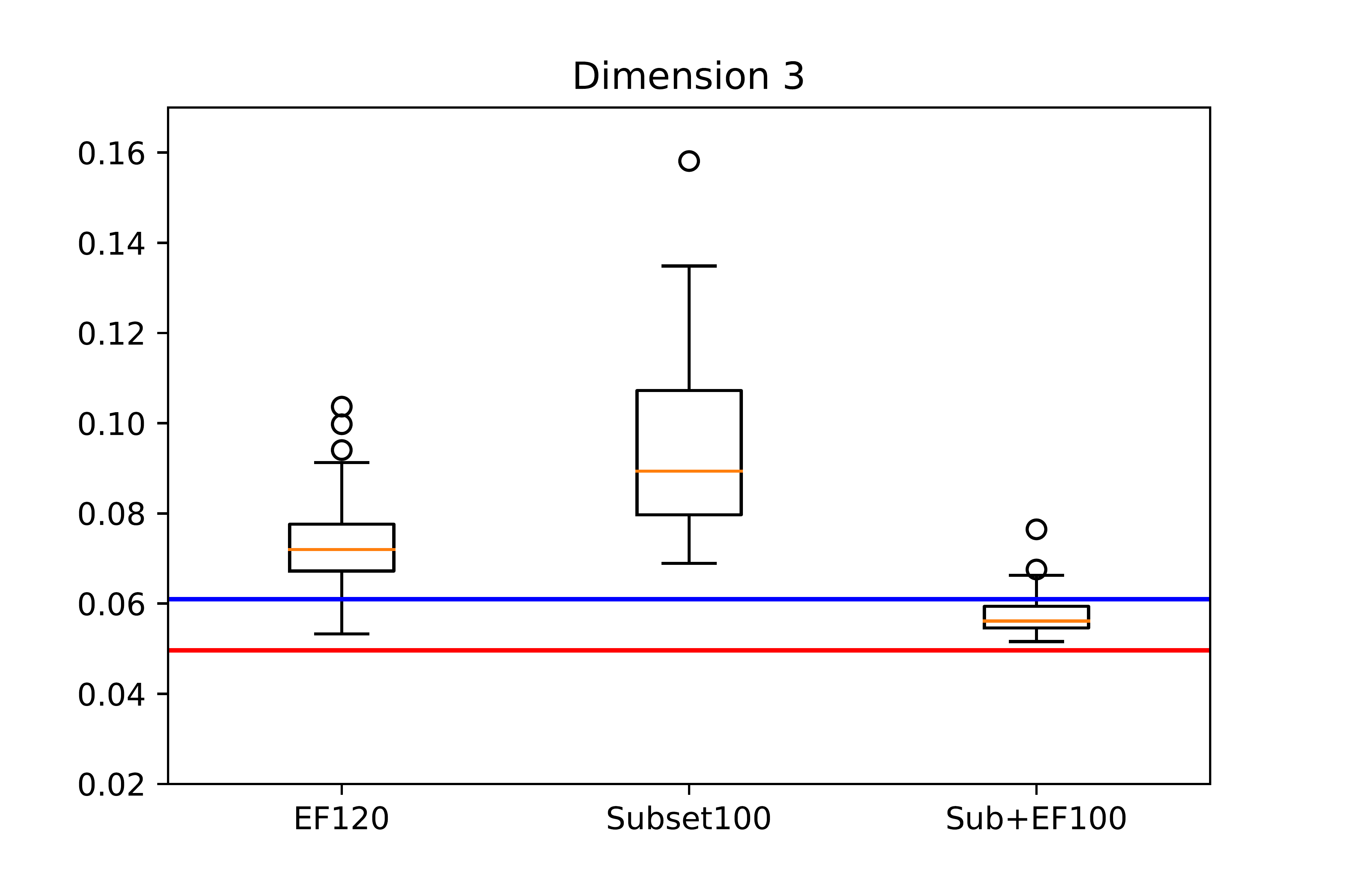}\hfill
    \includegraphics[width=.32\textwidth]{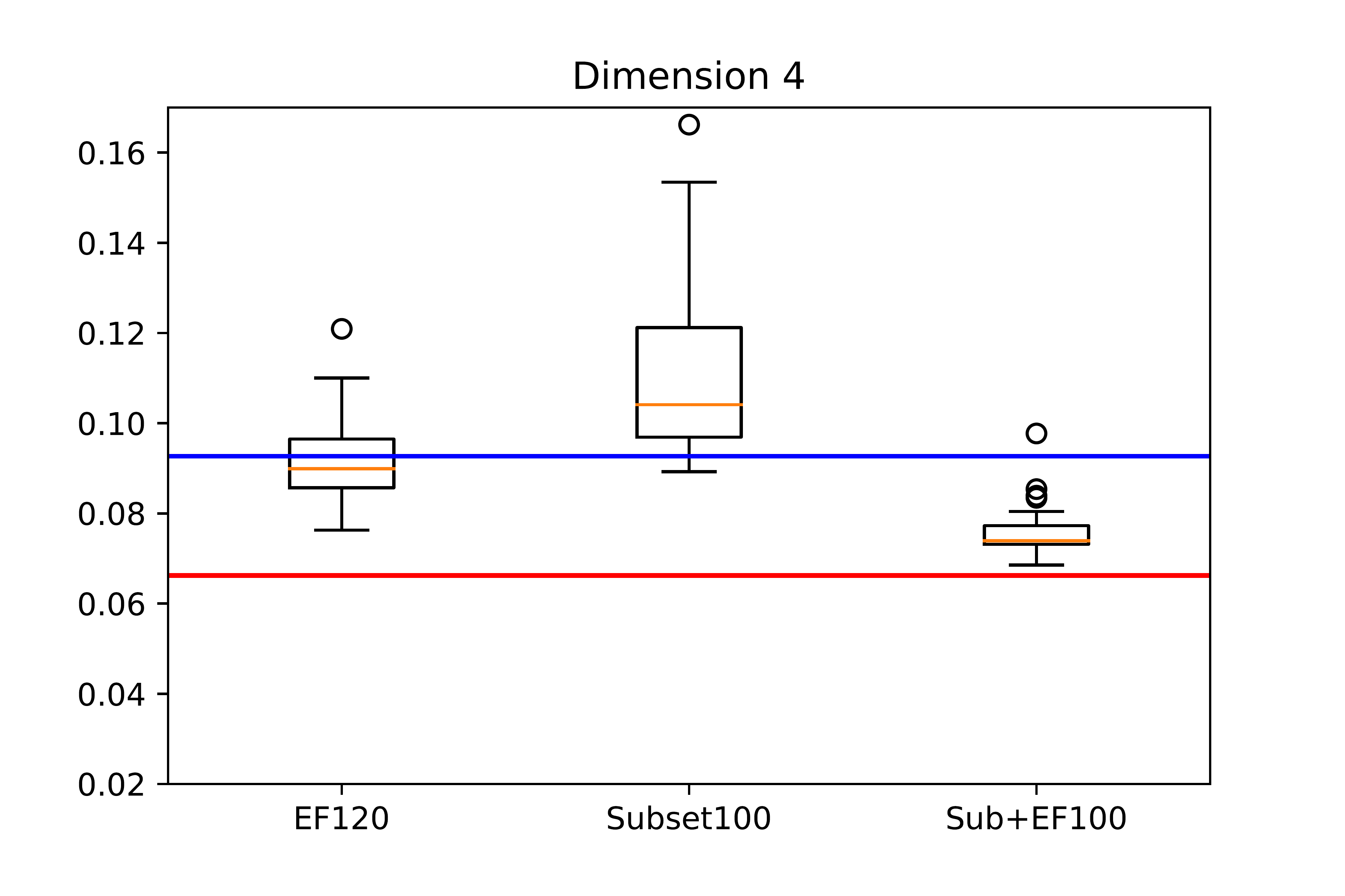}\hfill \smallskip
    \includegraphics[width=.32\textwidth]{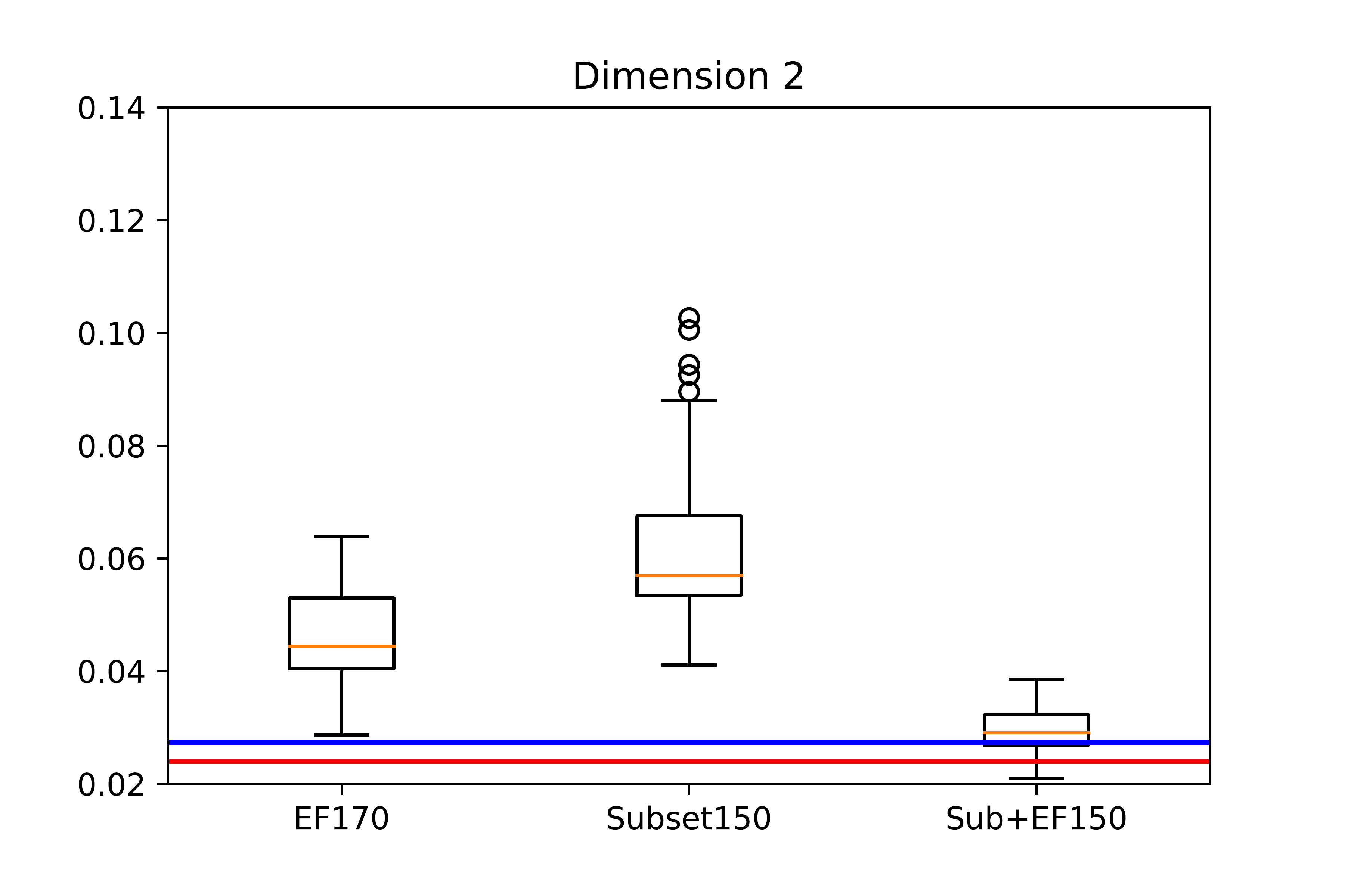}\hfill
    \includegraphics[width=.32\textwidth]{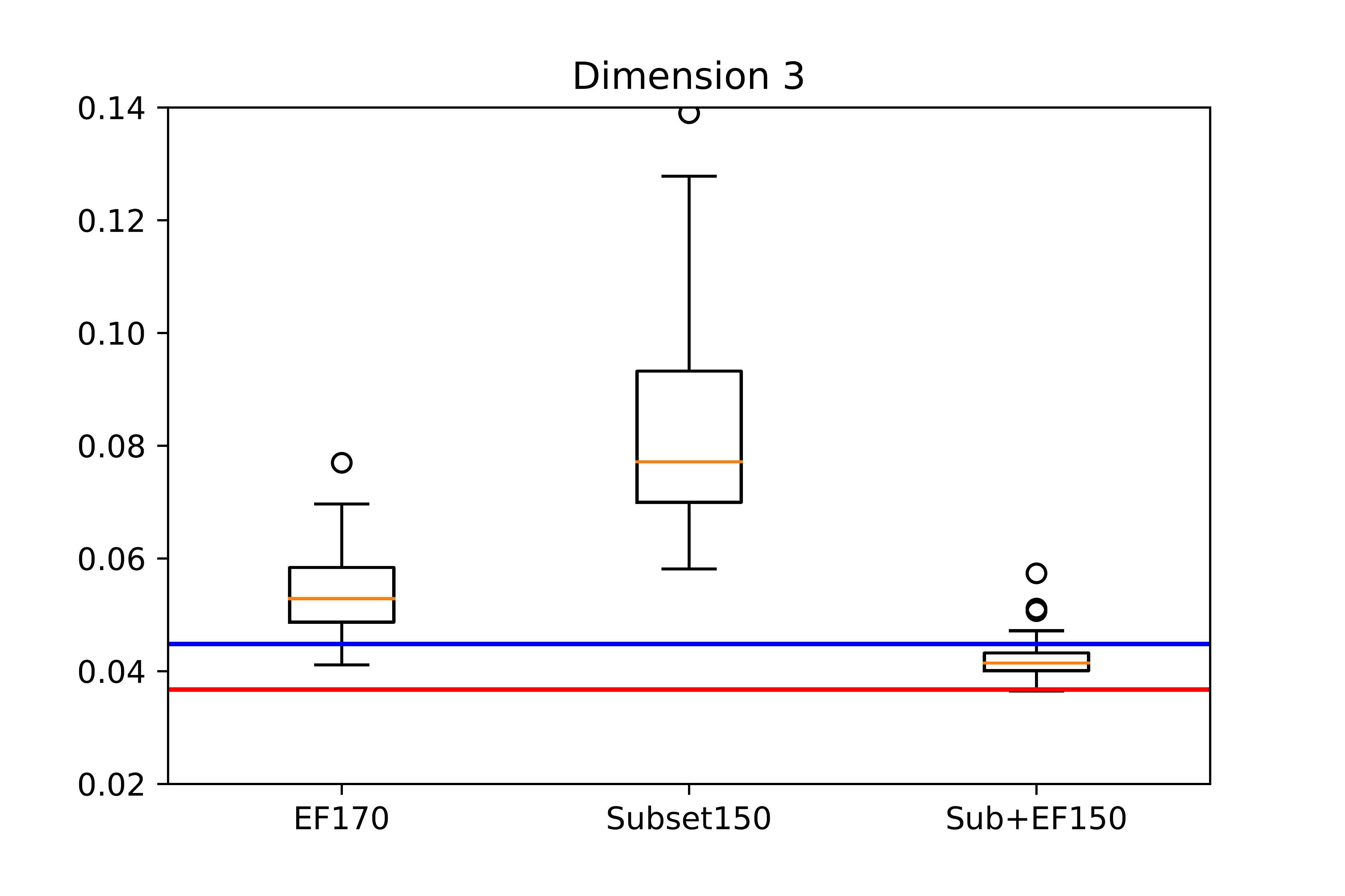}\hfill
    \includegraphics[width=.32\textwidth]{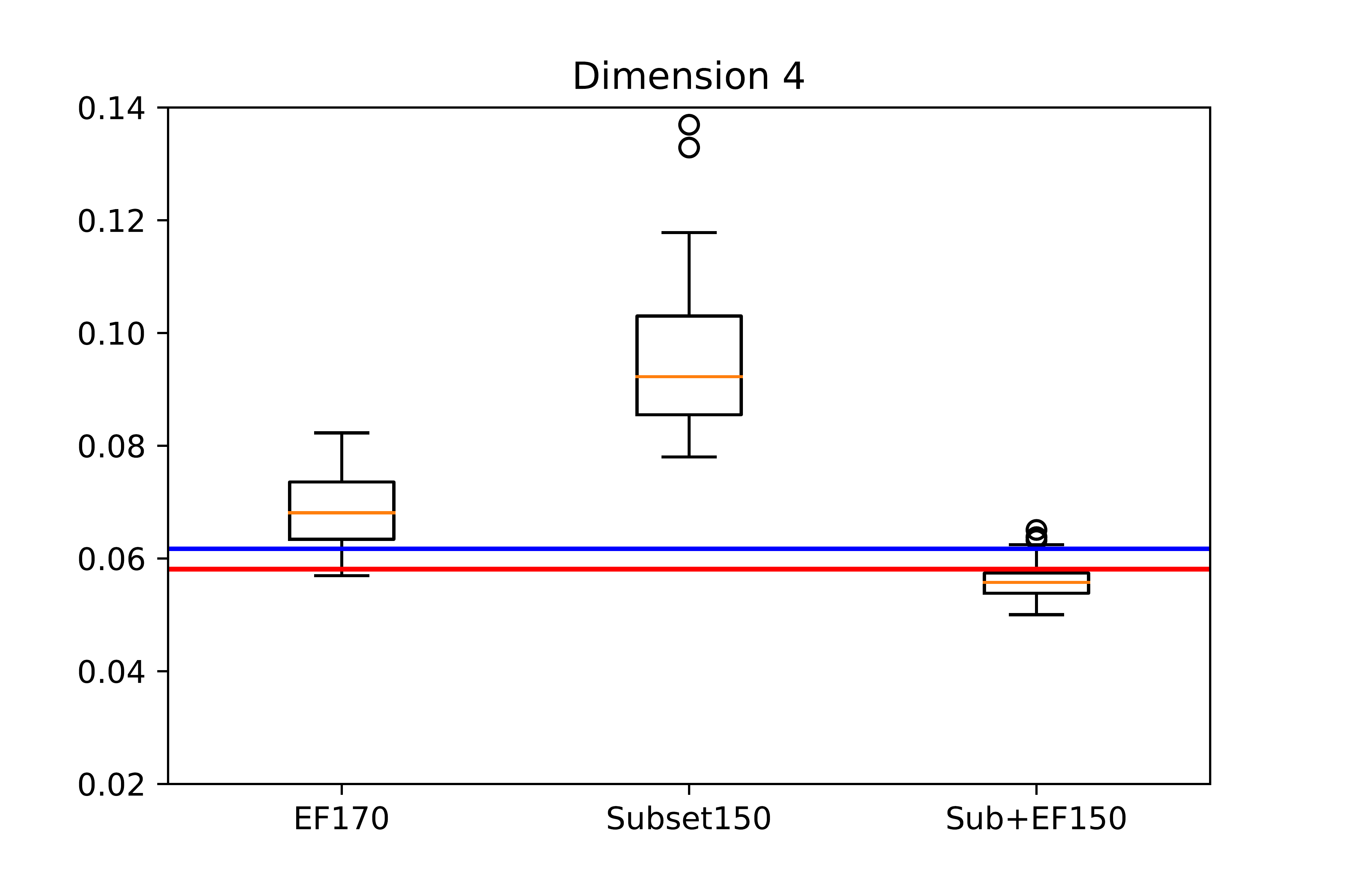}\hfill
    
    \caption{A comparison of subset selection (middle in each plot), the energy functional (left in each plot) and the combination of the two methods for random points (right in each plot). This is done in dimensions 2 (left), 3 (middle) and 4 (right) and with an initial $n=70$ (top), $n=120$ (middle) and $n=170$ (bottom). The horizontal lines represent the discrepancy values of the Sobol' sets of relevant size, $n$ (red) and $n-20$ (blue) in each plot.  }
    \label{CombiEner}
\end{figure}

\textbf{Impact of the ground set on the quality of the obtained point sets.} 
Despite this, the energy functional has one clear advantage over our method (apart from the runtime), in that it can take any point set as starting position. While this is not impossible for subset selection, the quality of the starting set strongly limits the quality of the resulting set with our method. For example, Figure~\ref{CombiEner} compares the effectiveness of subset selection, the energy functional and a combination of the two on random point sets generated in Python with the {\tt random} module. For each $(n,d)$ pair, 50 random instances are generated. The Sobol' sets are added as a comparison, the energy functional should be compared to the $n$ points line (red) and the two others to the $n-20$ line (blue). The sets obtained with only subset selection are a lot worse than the low-discrepancy sets, and in the majority of cases worse than those with only the energy functional. However, the combination of the two methods is always at least competitive with the Sobol' set of similar size, and even better in the vast majority of cases. This suggests a new method of computing low-discrepancy point sets, without requiring any knowledge of existing sequences or number theory: starting from any random set, applying successively the energy functional and then subset selection generates good low-discrepancy point sets. It also shows that while the discrepancy of the point sets obtained with the energy functional is not always as good as it could be, the point set created is regular enough to be used as a starting point for subset selection.

\section{Conclusion and Future Work}
Building on our previous paper on subset selection, we introduced a heuristic that allowed us to obtain better point sets in all dimensions for which the star discrepancy can be computed, with on average a 20\%  lower discrepancy than the initial point set. The obtained point sets were compared with known low-discrepancy sequences as well as with an energy functional by Steinerberger. We also provided some initial guidance on the optimal choice of parameters from the problem, with $k=n-20$ being a good baseline, where $k$ is the subset size and $n$ the input set size.

There are also a remaining number of open questions, both on our current heuristic and more general aspects of the problem. Firstly, we have only considered 1-swaps so far. It is likely that heuristics would perform better with more change possibilities at each step. The proposition in~\ref{Swaps} shows it would not bring better theoretical guarantees and it would entail more computations, but the heuristic would have a better capacity to explore the possible subsets. The second step is also very expensive: determining a limited set of pairs to check before initiating a restart rather than testing all combinations is likely to improve the heuristics. 

Secondly, this problem has shown the limits of current algorithms to compute the star discrepancy. They are expensive and can only give lower bounds when $n$ and $d$ get too high. Some sort of surrogate to replace the star discrepancy evaluations would be extremely useful, as well as interesting in itself to better understand the star discrepancy behavior. As we have shown, Steinerberger's functional would not be good enough for such purposes. A slightly less ambitious goal could be to find a better upper bound for the star discrepancy. Current upper bounds can be obtained either via Thiémard's approach~\cite{ThiBounds} or bracketing covers~\cite{GneBrack}, neither of which are fast enough for our purposes (despite recent improvements on the cover size in~\cite{GneBrackThie}). Improvements for these algorithms would lead to more precise information on the inverse star discrepancy. Our experiments already seem to show that known sequences perform better than expected, but faster and more precise algorithms would help us refine these conjectures.

\textbf{\textit{Acknowledgments.}} We thank Magnus Wahlstr\"om for providing his implementations of both the DEM algorithm and the TA heuristic. We also thank Stefan Steinerberger for providing his implementation in Mathematica of the energy functional and for interesting comments on it and on the inverse star discrepancy. We also thank Florian Pausinger for suggesting to compare our subset selection sets with low-discrepancy sets and not just sequences. Our work is financially supported by ANR-22-ERCS-0003-01 project VARIATION, by the CNRS INS2I project IOHprofiler, by Campus France Pessoa project 49173PH, and by national funds through the FCT - Foundation for Science and Technology, I.P. within the scope of the project CISUC – UID/CEC/00326/2020.

\appendix

\section{Existence of Local Minima}\label{App:heur}

\begin{proposition}\label{Swaps}
Let $k$-SDSSP-$j$ be the problem of obtaining the minimal star discrepancy subset of size $k$ by only doing improving $j$-swaps. For every $d \geq 2$, there exist point sets $P$ in dimension $d$ for which the $k$-SDSSP-$j$ has local minima which are not global minima if $n \geq 2k$ and $j <k$, or $n <2k$ and $j<n-k$.
\end{proposition}

\begin{figure}[h]
\centering
\begin{tikzpicture}
\draw  (4,0) -- (4,4) -- (0,4);
\draw[blue]  (0,3.333) -- (2.4,3.333) -- (2.4,0);
\draw[green]  (0,3.163) -- (2.4,3.163) -- (2.4,0);
\draw[red] [-stealth] (3.2,2.5) -- (3.2,2.22);
\draw[red] [-stealth] (3.6,2.22) -- (3.2,2.22);
\node[align=left] at (0.5,1) {Largest empty \\boxes for \\$P_A$ and $P_B$};
\node[align=left] at (5.3,3) {Top curve:\\ $xy=1- \alpha$};
\node[align=left] at (5.3,0.5) {Bottom curve:\\ $xy=1- 1/k$};

\draw (4 cm,1pt) -- (4 cm,-1pt) node[anchor=north] {$1$};
\draw (1pt,4 cm) -- (-1pt,4 cm) node[anchor=east] {$1$};
\filldraw[red, very thick] (1.97,4.0) -- (2.03,4.07) -- (2.08, 4.0) -- cycle;
\filldraw[red, very thick] (2.37,3.333) -- (2.43,3.403) -- (2.48,3.333) -- cycle;
\filldraw[red, very thick] (2.77,2.857) -- (2.83,2.927) -- (2.88,2.857) -- cycle;
\filldraw[red, very thick] (3.17,2.5) -- (3.23,2.57) -- (3.28, 2.5) -- cycle;
\filldraw[red, very thick] (3.57,2.222) -- (3.63,2.292) -- (3.68, 2.22) -- cycle;
\filldraw[red, very thick] (3.97,2.0) -- (4.03,2.07) -- (4.08,2.0) -- cycle;

\filldraw[blue, very thick] (2.0,3.333) rectangle (2.06,3.393);
\filldraw[blue, very thick] (2.4,2.857) rectangle (2.46,2.917);
\filldraw[blue, very thick] (2.8,2.5) rectangle (2.86,2.56);
\filldraw[blue, very thick] (3.2,2.222) rectangle (3.26,2.282);
\filldraw[blue, very thick] (3.6,2.0) rectangle (3.66,2.06);

\foreach \Point in {(1.8,3.163), (2.4,2.687), (2.8,2.33), (3.2,2.052), (3.6,1.83)}{
    \node[green] at \Point {\textbullet};

}

\draw plot [smooth] coordinates {(1.6,5.0) (2.0,4.0) (2.4,3.333) (2.8,2.857) (3.2,2.5) (3.6,2.222) (4.0,2.0) (5.0,1.6)};

\draw plot [smooth] coordinates {(1.2,4.583) (1.375,4) (1.6,3.4375) (2.0,2.75) (2.4,2.291) (2.8,1.964) (3.2,1.719) (3.6,1.528) (4.0,1.375) (4.583,1.2)};

\end{tikzpicture}
\caption{An illustration of the different points of the proof of Proposition~\ref{Swaps}: the $q_i$ are in red, $p_i$ in blue if $i<k+1$ and green otherwise. The lower curve corresponds to $xy=1-1/k$, the upper one to $xy=1-\alpha$. They are not up to scale for readability. The red lines represent how blue points are built, whereas the blue and green boxes are the discrepancy-defining boxes for $P_A$ and $P_B$ respectively. The $L_{\infty}$ star discrepancy of $P_A$ is $1-\alpha$ while that of $P_B$ is $1-\alpha-\delta p_{2,1}$. However, it is impossible to transition from $P_A$ to $P_B$ without changing the whole set at once.}
\label{illus}
\end{figure}
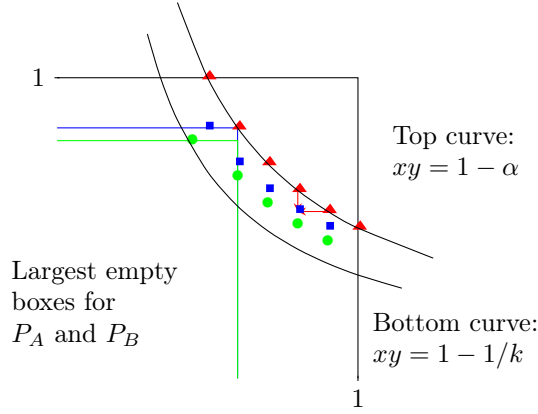

\begin{proof}
We first consider a base case for $j=k-1$, $d=2$, and $n=2k$. Its extension to all the other combinations of $d$, $j$, and $n$ will be described afterwards. 

\textbf{Base case:} We construct a point set $P\subseteq [0,1]^2$ of size $2k$ and let $j=k-1$, represented in Figure~\ref{illus}. To build $P$, we first consider a very small constant $\alpha$ and a set of $k+1$ points $(q_{i,1},q_{i,2})$ that satisfy $q_{i,1}q_{i,2}=1-\alpha$. We assume the indices to be sorted such that for all $i$, $q_{i,1} < q_{i+1,1}$. We further require that $q_{i,1}q_{i+1,2} > 1-1/(2k)$. These points are the red triangles in Figure~\ref{illus}. We then use these $k+1$ points to build the first $k$ points of $P$, the set $P_A:=\{(q_{i,1},q_{i+1,2}): 1\leq i \leq k\}:=\{(p_{i,1},p_{i,2}):i \in \{1,\ldots,k\}\}$ given by blue squares in Figure~\ref{illus}. From each point $p_i$ in $P_A$ with $i \geq 2$, we can build a point $p_{k+i}$, with $p_{k+i,1}:=p_{i,1}$ and $p_{k+i,2}:= p_{i,2}-\delta$, where $\delta$ is a small positive constant strictly upper-bounded by $\min_{i\in \{1,\ldots,k-1\}}|p_{i,2}-p_{i+1,2}|$ and such that $p_{k+i,1}p_{k+i,2} \geq 1-1/k$. $p_{k+1}$ is such that $p_{k+1,2}=p_{1,2}- \delta$ and $p_{k+1,1}=p_{1,1}-\gamma$, where $\gamma$ is strictly positive smaller than $p_{1,1}$ and such that $p_{k+1,1}p_{k+1,2} \geq 1-1/k$. The set formed by these points is defined as $P_B$, represented by green discs in Figure~\ref{illus}.

We first note that for any subset $P_k$ of $P$ of size $k$, the $L_{\infty}$ star discrepancy of $P_k$ will be given by the largest box in $[0,1]^d$ containing no points of $P_k$. Any open box containing points will have local discrepancy at most $1-1/k$ (the maximal volume minus the minimal number of points) and any closed box containing points will have local discrepancy at most $1- (1-1/k)=1/k$ (the maximal number of points minus the minimal volume of a box containing a point). On the other hand, the largest empty box will always have volume at least $1-1/k$ (all the points are above the curve $xy=1-1/k$, see Figure~\ref{illus}) and thus local discrepancy at least $1-1/k$.

We now show that $P_A$ is a local optimum but not a global one. First of all, $P_A$ has discrepancy exactly $1-\alpha$, obtained for one of the $k+1$ empty boxes whose top-right corner is either $(p_{1,1},1)$, $(1,p_{k,2})$ or $(p_{i+1,1},p_{i,2})$ for $i \in \{1,\ldots,k-1\}$. By definition of $j$-swaps, we replace $j$ of the points inside $P_A$ by points in $P_B$ (since $P_A \cup P_B=P$). Let $P_C$ be our new point set,  $|P_A \cap P_C|=1$. Let $p_i$ be the point in $P_A \cap P_C$. There are now two different cases to consider:
\begin{itemize}
\item If $p_{k+i}$ is not in $P_C$, then the box with upper-right corner in $(p_{i+1,1},p_{i,2})$ is still empty. None of the other points could be inside since we have $p_{1,2} \geq p_{k+1,2} \geq p_{2,2} \geq \ldots \geq p_{k,2} \geq p_{2k,2}$ and the first coordinates are ordered in the reverse order. The discrepancy of $P_C$ is therefore at least that of $P_A$.
\item If $p_{k+i}$ is in $P_C$, then there exists $h \in \{1,\ldots,k\}$ such that $p_h$ and $p_{k+h}$ are not in $P_C$. The box with upper-right corner in $(p_{h+1,1},p_{h-1,2})$ (with $p_{h+1,1}=1$ if $h+1 >k$ and $p_{h-1,2}=1$ if $h=1$) has to be empty. By the ordering given above, this  empty box will have a volume greater than that of $p_{h+1,1}p_{h,2}$, the discrepancy of $P_A$.

\end{itemize}
By the above, $P_A$ is a local minimum. However, for $P_B$, the largest empty boxes will have volume $p_{k+i+1,1}p_{k+i,2}$ for $i \in \{1,\ldots,k-1\}$. By construction $p_{k+i+1,1}=p_{i+1,1}$ and $p_{k+i,2}<p_{i,2}$. Since the discrepancy of $P_B$ is given by the largest empty box, we can conclude that $d_{\infty}^{*}(P_B)<d_{\infty}^{*}(P_A)$, $P_A$ is hence a local minimum for the $k$-SDSSP-$j$ problem, but not a global one.

\textbf{Higher dimensions:} Taking the same point set with $1$'s added for all the coordinates in dimensions greater than 2 gives the exact same proof.

\textbf{More points $n>2k$:} We can add all the $n-2k$ new points to the region above the curve $xy=1-\alpha$. Taking any of these never reduces the volume of the largest empty box, thus a local/global minimum in the base case is still a local/global optimum.

\textbf{Fewer swaps $j<k-2$:} A local optimum for $j=k-2$ swaps is still a local optimum for $j<k-2$ swaps, the global optimum is unchanged.

\textbf{Fewer points $n<2k$:} The same construction is no longer possible and at least $2k-n$ points have to be shared between any two sets. If $j \geq n-k$ then there are no local optima which are not also global optima, as any subset can be transformed to any other in a single step. If we remove the first $2k-n$ points from $P$ (this is less than $n$, we are removing only points from $P_A$), we want the set $P_{LO}:=\{p_i: i \in \{2k-n+1,3k-n\}\}$ to be a local minimum. We note that we are keeping the numbering from the base case, i.e, the points in $P$ are numbered from $2k-n+1$ to $2k$. This requires us to be unable to switch all the $n-k$ ``unshared'' points to those in $P_B$ (note that $p_1$ and $p_{k+1}$ no longer have a special role as $p_1$ no longer exists), therefore $j< n-k$. The proof is then the same. 
\end{proof}
\newpage
\section{Computational Results}
\label{App}
We include in this Appendix some of the discrepancy values obtained with the different methods. Unless they are in \textbf{bold}, values from TA sets were verified with the DEM algorithm. This could lead to higher discrepancy values than those obtained during the subset selection heuristic, and possibly that this heuristic missed better sets because of those mistakes. The corrected values for the TA heuristics only include those from \emph{concluded} runs. A large number of the bold values simply correspond to sets where the best run was interrupted and the stored set was clearly poorer. 
\begin{center}
\begin{table}[h!]
\centering

\caption{Discrepancy values obtained in dimension 4 for the different heuristics, the DEM\_NBF version was not run for $n \geq 200$.}
\begin{tabular} {|c c  c  c  c c|}
\hline
Set size & Subset size $k$ & DEM\_BF & DEM\_NBF & TA\_BF & TA\_NBF \\
\hline
$n=50$ & $k=50$ & 0.13422 & 0.13422 & 0.13422 & 0.13422 \\

 & 40 & 0.12236 & 0.12520 & \textbf{0.13406} & 0.14090 \\
 & 30 & 0.14020 & 0.14924 & \textbf{0.15431} & 0.15471 \\
 & 20 & 0.17660 & 0.18494 & \textbf{0.17883} & 0.20066 \\

 \hline
 $n=100$ & $k=100$ & 0.092688 & 0.092688  & 0.092688 & 0.092688 \\
 & 90 & 0.070093 & 0.075315 & \textbf{0.076933} & 0.075315 \\
 & 80 & 0.071985 & 0.082731 & \textbf{0.084182} & 0.08625\\
 & 70 & 0.078701 & 0.087342 & 0.09545 & 0.088841 \\
 & 60 & 0.087650 & 0.095528 & \textbf{0.103849} & 0.100801 \\
 & 50 & 0.097189 & 0.110063 & \textbf{0.119433} & 0.118624 \\
    
\hline

 $n=150$ & $k=150$ & 0.061738 &0.061738  & 0.061738 & 0.061738 \\
 & 140 & 0.052081 & 0.054116 & 0.056260 & \textbf{0.054268} \\
 & 130 & 0.051702 & 0.057176 & 0.060173 & 0.060173\\
 & 120 & 0.056405 & 0.062694 & 0.065592 & 0.065592 \\
 & 110 & 0.059261 & 0.063807 & \textbf{0.067161} & 0.070707 \\
 & 100 & 0.061478 & 0.068499 & \textbf{0.077275} & \textbf{0.077275} \\

\hline

 $n=200$ & $k=200$ & - & 0.050215  & 0.050215 & 0.050215 \\
 & 190 & - & 0.045960 & 0.054374 & 0.045960 \\
 & 180 & - & 0.046837 & - & 0.046848\\
 & 170 & - & 0.048267 & 0.054912 & 0.052956 \\
 & 160 & - & 0.051588 & 0.065464 & 0.054244 \\
 & 150 & - & 0.053195 & - & \textbf{0.055839} \\

\hline

 $n=250$ & $k=250$ & - & 0.038216  & 0.038216 & 0.038216 \\
 & 240 & - & 0.036286 & 0.040657 & 0.037994 \\
 & 230 & - & 0.037675 & 0.040731 & 0.040603\\
 & 220 & - & 0.037972 & - & \textbf{0.043796} \\
 & 210 & - & 0.039900 & - & 0.042615 \\
 & 200 & - & 0.043015 & - & 0.046989 \\
 \hline

 $n=500$ & $k=500$ & - & 0.022901  & - & 0.022901 \\
 & 490 & - & 0.021662 & - & \textbf{0.021187} \\
 & 480 & - & 0.021572 & - & \textbf{0.022823}\\
 & 470 & - & 0.021211 & - & 0.024016 \\
 & 460 & - & 0.022744 & - & 0.025268 \\
 & 450 & - & 0.023916 & - & 0.027314 \\

 \hline

\end{tabular}
\label{App_Dim4}

\end{table}

\end{center}
\newpage

\begin{center}
\begin{table}[h!]
\centering
\caption{Discrepancy values obtained in dimension 5 for the different heuristics, the DEM\_NBF version was not run for $n \geq 200$. The - in the DEM column indicate that not a single run finished within the time limit.}
\begin{tabular} {|c c  c  c  c c|}
\hline
Set size & Subset size $k$ & DEM\_BF & DEM\_NBF & TA\_BF & TA\_NBF \\
\hline
$n=50$ & $k=50$ & 0.165488 & 0.165488 & 0.165488 & 0.165488 \\

 & 40 & 0.138741 & 0.149003 & \textbf{0.148836} & 0.147256 \\
 & 30 & 0.161715 & 0.171163 & 0.17149 & 0.17149 \\
 & 20 & 0.211900 & 0.214233 & 0.234375 & 0.23438 \\

 \hline
 $n=100$ & $k=100$ & 0.120707 & 0.120707  & 0.120707 & 0.120707 \\
 & 90 & 0.086374 & 0.092956 & \textbf{0.090522} & 0.091191 \\
 & 80 & 0.090923 & 0.095958 & \textbf{0.099937} & \textbf{0.097624}\\
 & 70 & 0.099563 & 0.105703 & \textbf{0.109832} & 0.116787 \\
 & 60 & 0.107044 & 0.117161 & 0.125410 & \textbf{0.118965} \\
 & 50 & 0.118428 & 0.129599 & \textbf{0.135023} & \textbf{0.134716} \\
    
\hline

 $n=150$ & $k=150$ & 0.074899 & 0.074899  & 0.074899 & 0.074899 \\
 & 140 & 0.064423 & 0.054116 & \textbf{0.054163} & \textbf{0.054268} \\
 & 130 & 0.064549 & 0.057176 & \textbf{0.060173} & \textbf{0.060173}\\
 & 120 & 0.068790 & 0.062694 & \textbf{0.063046} & \textbf{0.063046} \\
 & 110 & 0.073173 & 0.063807 & \textbf{0.067161} & \textbf{0.067161} \\
 & 100 & 0.077755 & 0.068499 & \textbf{0.077275} & \textbf{0.077275} \\

\hline

 $n=200$ & $k=200$ & - & 0.058292  & 0.058292 & 0.058292 \\
 & 190 & - & 0.053908 & 0.059209 & \textbf{0.053496} \\
 & 180 & - & 0.055425 & 0.061967 & \textbf{0.058496}\\
 & 170 & - & 0.059826 & 0.067392 & 0.062506 \\
 & 160 & - & 0.061195 & 0.075936 & 0.065796 \\
 & 150 & - & 0.064438 & - & \textbf{0.067725} \\

\hline

 $n=250$ & $k=250$ & - & 0.053507  & 0.053507 & 0.053507 \\
 & 240 & - & 0.044187 & 0.048002 & \textbf{0.048310} \\
 & 230 & - & 0.046281 & 0.046463 & 0.046463\\
 & 220 & - & 0.048236 & 0.053760 & 0.052865 \\
 & 210 & - & 0.049783 & - & \textbf{0.055719} \\
 & 200 & - & 0.052454 & 0.062576 & \textbf{0.055284} \\
 \hline

 $n=500$ & $k=500$ & - & 0.029117  & - & 0.029117 \\
 & 490 & - & 0.028160 & - & 0.029485 \\
 & 480 & - & 0.029255 & - & 0.030663\\
 & 470 & - & 0.030951 & - & 0.031702 \\
 & 460 & - & - & - & 0.034878 \\
 & 450 & - & - & - & 0.036084 \\

 \hline

\end{tabular}
\label{App_Dim5}

\end{table}

\end{center}
\newpage

\begin{center}
\begin{table}[h!]
\centering
\caption{Discrepancy values obtained in dimension 6 for the different heuristics, the DEM\_NBF version was not run for $n \geq 200$. Other - correspond to unfinished runs.}
\begin{tabular} {|c c  c  c  c c|}
\hline
Set size & Subset size $k$ & DEM\_BF & DEM\_NBF & TA\_BF & TA\_NBF \\
\hline
$n=50$ & $k=50$ & 0.225548 & 0.225548 & 0.225548 & 0.225548 \\

 & 40 & 0.162600 & 0.166336 & \textbf{0.166522} & 0.169385 \\
 & 30 & 0.186518 & 0.197138 & \textbf{0.197326} & 0.194999 \\
 & 20 & 0.232082 & 0.246097 & 0.26041 & 0.253279 \\

 \hline
 $n=100$ & $k=100$ & 0.124451 & 0.124451  & 0.124451 & 0.124451 \\
 & 90 & 0.100532 & 0.102214 & \textbf{0.100660} & 0.11364 \\
 & 80 & 0.109108 & 0.114143 & \textbf{0.113169} & \textbf{0.114344}\\
 & 70 & 0.120823 & 0.120817 & 0.133813 & 0.120032 \\
 & 60 & 0.128823 & 0.134782 & \textbf{0.140478} & \textbf{0.131683} \\
 & 50 & 0.139858 & 0.149740 & \textbf{0.161732} & \textbf{0.150491} \\
    
\hline

 $n=150$ & $k=150$ & 0.090827 & 0.090827  & 0.090827 & 0.090827 \\
 & 140 & - & 0.081020 & \textbf{0.078646} & \textbf{0.082801} \\
 & 130 & - & 0.085759 & \textbf{0.078646 }  & \textbf{0.084272}\\
 & 120 & - & 0.089289 & 0.091492 & \textbf{0.090094} \\
 & 110 & - & 0.090785 & 0.103782 & \textbf{0.097381} \\
 & 100 & - & 0.100891 & 0.104589 & 0.107580 \\

\hline

 $n=200$ & $k=200$ & - & 0.087784 & - & 0.087784 \\
 & 190 & - & 0.068581 & 0.078203 & \textbf{0.065424} \\
 & 180 & - & 0.070267 & 0.077221 & \textbf{0.065122}\\
 & 170 & - & - & 0.081820 & \textbf{0.071578} \\
 & 160 & - & 0.077993 & - & \textbf{0.072786} \\
 & 150 & - & 0.084923 & - & \textbf{0.079571} \\

\hline

 $n=250$ & $k=250$ & - & 0.088941  & - & 0.088941 \\
 & 240 & - & 0.064764 & - & \textbf{0.057167} \\
 & 230 & - & - & - & \textbf{0.060105}\\
 & 220 & - & - & - & \textbf{0.064285} \\
 & 210 & - & - & - & \textbf{0.062227} \\
 & 200 & - & 0.073417 & - & \textbf{0.068424} \\
 \hline

 $n=500$ & $k=500$ & - & 0.040529  & - & 0.040529 \\
 & 490 & - & - & - & 0.036165 \\
 & 480 & - & - & - & \textbf{0.034287}\\
 & 470 & - & - & - & \textbf{0.038244} \\
 & 460 & - & - & - & 0.042268 \\
 & 450 & - & - & - & 0.035382 \\

 \hline

\end{tabular}
\label{App_Dim6}

\end{table}

\end{center}
\newpage

\begin{center}
\begin{table}[h!]
\centering
\caption{Discrepancy values obtained in dimension 8, 10, 15 and 25 for TA. DEM\_NBF could finish only for the smallest $k$ in dimension 8 and is omitted here. \textbf{None of the values were verified with the exact algorithm}. }
\begin{tabular} {|c c  c c c c |}
\hline
Set size & Subset size $k$ &TA\_NBF $d=8$ & TA\_NBF $d=10$ & TA\_NBF $d=15$ & TA\_NBF $d=25$ \\
\hline
$n=50$ & $k=50$ & 0.248547 & 0.298001 & 0.388598 & 0.483923\\

 & 40 & 0.207029 & 0.293720 & 0.340869 & 0.459241\\
 & 30 & 0.236120 & 0.270010 & 0.342629 & 0.459946\\
 & 20 & 0.293341 & 0.339852 & 0.404245 & 0.522859\\

 \hline
 $n=100$ & $k=100$ & 0.160793 & 0.208052 & 0.258440 & 0.339362\\
 & 90 & 0.137759 & 0.167185 & 0.233544 & 0.316399\\
 & 80 & 0.140670 & 0.168474 & 0.232660 & 0.316893\\
 & 70 & 0.147338 & 0.177223 & 0.238510 & 0.329086\\
 & 60 & 0.156791 & 0.193031 & 0.253804 & 0.342430\\
 & 50 & 0.173767 & 0.209863 & 0.274870 & 0.367155\\
    
\hline

 $n=150$ & $k=150$ & 0.106714 & 0.150029 & 0.193065 & 0.273853\\
 & 140 & 0.100294 & 0.126767 & 0.175397 & 0.251457\\
 & 130 & 0.099135 & 0.124570 & 0.175082 & 0.255210\\
 & 120 & 0.106236 & 0.136273 & 0.177053 & 0.250421\\
 & 110 & 0.116722 & 0.142494 & 0.190030 & 0.263972\\
 & 100 & 0.125779 & 0.146189 & 0.204310 & 0.282990\\

\hline

 $n=200$ & $k=200$ & 0.095888 & 0.120527 & 0.167833 & 0.228047\\
 & 190 &0.085177 & 0.107770 & 0.150436 & 0.218568\\
 & 180 & 0.084830 & 0.107667 & 0.145002 & 0.215581\\
 & 170 & 0.091665 & 0.118012 & 0.150607 & 0.208230\\
 & 160 & 0.097845 & 0.110122 & 0.157870 & 0.220264\\
 & 150 & 0.095667 & 0.124148 & 0.163522 & 0.229887\\

\hline

 $n=250$ & $k=250$ & 0.078968 & 0.097270 & 0.163522 & 0.207971\\
 & 240 & 0.070861 & 0.089403 & 0.141638 & 0.194531\\
 & 230 & 0.076617 & 0.094512 & 0.126737 & 0.192541\\
 & 220 & 0.076390 & 0.096894 & 0.127134 & 0.194215\\
 & 210 & 0.084471 & 0.098762 & 0.129550 & 0.198693\\
 & 200 & 0.079550 & 0.105629 & 0.144049 & 0.201951\\
 \hline

 $n=500$ & $k=500$ & 0.047839 & 0.061573 & 0.086172 & 0.147068\\
 & 490 & 0.047199 & 0.060665 & 0.082243 & 0.135416\\
 & 480 & 0.046720 & 0.061057 & - & -\\
 & 470 & 0.049966 & - & - & -\\
 & 460 & - & 0.063294 & - & -\\
 & 450 & 0.053917 & 0.066657 & - & - \\

 \hline

\end{tabular}
\label{App_Dim8}

\end{table}

\end{center}

\newpage

\bibliographystyle{elsarticle-harv}
\bibliography{references}

\end{document}